\documentclass[journal]{IEEEtran}

\usepackage{graphicx}
\usepackage{float}
\usepackage{subfigure}
\usepackage{amssymb}
\usepackage{amsmath}
\usepackage{braket}
\usepackage{comment}
\usepackage{url}
\usepackage{multirow}
\usepackage{booktabs}
\begin{document}

\title{Adaptive Hamiltonian Identification via Quantum Reinforcement Learning under Non-Stationary Dynamics}

\author{\IEEEauthorblockN{Jawaher Kaldari\IEEEauthorrefmark{1}, Saif Al-Kuwari\IEEEauthorrefmark{1}}

\IEEEauthorblockA{\IEEEauthorrefmark{1} Qatar Center for Quantum Computing, College of Science and Engineering, Hamad Bin Khalifa University, Doha, Qatar }

Emails: \{jaka51804, smalkuwari\}@hbku.edu.qa}

\maketitle

\begin{abstract}

Accurate identification of the Hamiltonian governing a quantum system is important for quantum-system characterization, device calibration, simulation, and reliable operation of quantum hardware. In practical systems, however, the governing dynamics may vary over time because of parameter drift, changing control conditions, or transitions between operating regimes. In this work, we formulate Hamiltonian identification as a quantum reinforcement learning (QRL) problem in which a quantum agent processes sequential quantum-state observations and estimates the governing Hamiltonian parameters. We first consider stationary dynamics using a Vanilla-QRL agent and then introduce a recurrent GRU-QRL architecture to track non-stationary dynamics. The framework is evaluated on transverse-field Ising model systems under stationary dynamics, abrupt Hamiltonian switching, rapidly changing Hamiltonians, random switch locations over longer trajectories, and gradual Hamiltonian drift. We further compare the proposed approach with a Quantum Graph Recurrent Neural Network (QGRNN) baseline and evaluate frozen policies on unseen trajectories. Results show that both approaches accurately reconstruct stationary Hamiltonians, while GRU-QRL provides substantially improved tracking under non-stationary dynamics. These results demonstrate the potential of recurrent QRL as an adaptive system-identification tool for monitoring time-varying quantum systems.

\end{abstract}


\IEEEpeerreviewmaketitle

\section{Introduction}

\IEEEPARstart{Q}{uantum} technologies are rapidly evolving from small-scale proof-of-concept devices to large-scale systems capable of tackling problems beyond the reach of classical supercomputers. As these systems continue to grow in scale and complexity, verifying, calibrating, and controlling their behavior becomes increasingly challenging, motivating the need for automated, data-driven methods to characterize the underlying quantum dynamics. Hamiltonian learning has recently emerged as a key tool for system identification \cite{burgarth2012quantum}, device calibration \cite{cole2005identifying}, and error correction \cite{valenti2019hamiltonian}, all of which are essential for the reliable operation of quantum devices. Hamiltonian learning aims to infer the underlying Hamiltonian governing a quantum system from experimentally accessible data, such as measurement outcomes, quantum state trajectories, or local observables, providing deep structural insights into the system's physical behavior.

Beyond system identification and device calibration, Hamiltonian learning also plays a fundamental role in quantum simulation \cite{guo2025hamiltonian}. As quantum simulators continue to scale and are increasingly used to investigate complex many-body systems, verifying that the implemented dynamics follow the intended Hamiltonian becomes a significant challenge. An early straightforward approach is quantum process tomography (QPT), which reconstructs the complete quantum process to verify the system's evolution \cite{lanyon2011universal}. However, the computational and experimental resources required for QPT grow exponentially with the system size, making it impractical for large-scale quantum systems \cite{zhao2025learning}. Hamiltonian learning provides a more efficient alternative by estimating the underlying Hamiltonian directly from experimental observations.

From an engineering perspective, Hamiltonian identification is particularly important when quantum systems operate over extended periods and their effective dynamics cannot be assumed to remain perfectly stationary. Changes in control fields, device parameters, calibration conditions, or environmental interactions can modify the effective Hamiltonian and consequently degrade the accuracy of previously calibrated system models. An adaptive identification method capable of tracking such changes from sequential observations could therefore support continuous characterization, recalibration, and monitoring of quantum devices. This motivates the non-stationary Hamiltonian-learning setting considered in this work.

A wide range of Hamiltonian learning algorithms have been proposed. Existing approaches differ in the assumptions they make about the available experimental data and the underlying physical system, as well as in the learning methodologies they employ. 

Early Hamiltonian learning methods were primarily based on Bayesian framework. In \cite{granade2012robust}, an adaptive Bayesian approach was proposed, combining Monte Carlo with Bayesian experimental design, in which each subsequent experimental control is chosen based on the outcomes observed so far. The method takes as input a sequence of binary measurement outcomes obtained from repeated experimental runs, and outputs an estimate of the underlying Hamiltonian parameters along with a region estimate quantifying the uncertainty in that estimate. A separate line of work instead formulates Hamiltonian learning as solving a system of linear equations \cite{qi2019determining,bairey2019learning}. A more recent work, \cite{evans2019scalable}, introduces a Bayesian formulation to this linear-system approach, achieving greater scalability and robustness to noise, drawing on the sequential updating idea from \cite{granade2012robust}.

Distinct from both the Bayesian and linear-system approaches, the authors in \cite{gu2024practical} proposed a Hamiltonian learning method for many-body quantum systems. Their approach estimates the unknown Hamiltonian coefficients from local measurements of the system dynamics using derivative estimation and Chebyshev polynomial regression. The method was demonstrated on the transverse-field Ising model and showed scalability to large quantum systems

More recently, machine learning techniques have been applied to the Hamiltonian learning problem. The authors in \cite{heightman2025solving} proposed a neural differential equation-based approach to infer many-body Hamiltonians from quantum state trajectories. Similarly, in \cite{cemin2024inferring}, machine learning was used to infer the effective dynamical generator of a quantum subsystem from noisy projective measurements. Their method learns a physically interpretable Lindblad generator, allowing the underlying Hamiltonian parameters and decoherence processes to be extracted from the learned model. Other machine learning methods use a convolutional conditional generative adversarial network to estimate Hamiltonian parameters from simulated and experimental differential conductance measurements \cite{koch2023adversarial}. In \cite{an2025dual}, LSTM-based supervised learning was used to infer time-dependent Hamiltonian parameters from the evolution of local observables and demonstrated its performance under different driving fields, including quenches and periodic driving.

Building on the growing use of machine learning for Hamiltonian learning, advances in quantum machine learning have motivated the exploration of quantum learning approaches for this task. In \cite{verdon2019quantum}, Quantum Graph Recurrent Neural Network (QGRNN) was applied to Hamiltonian learning: given copies of a fixed initial state and its time-evolved at various times, the QGRNN is trained via gradient descent to learn the couplings and biases of a graph-structured Ising Hamiltonian with a transverse field. Moreover, variational quantum learning was utilized for Hamiltonian learning using time-resolved measurement data \cite{sohail2026quantum}. Their method learns the unknown Hamiltonian parameters by minimizing negative log-likelihood loss over measurement outcomes collected at different evolution times. In addition, variational quantum circuits were combined with classical energy-based models to learn a Hamiltonian representation of the underlying data distribution for generative modeling and anomaly detection \cite{araz2023quantum}.

In this work, we present a quantum reinforcement learning (QRL) framework that infers the parameters of a Hamiltonian governing a closed quantum system from a sequence of time-evolved quantum states. The proposed framework formulates Hamiltonian learning as an autonomous sequential decision-making problem within a QRL framework. We further investigate the proposed framework under non-stationary quantum dynamics, where the governing Hamiltonian changes during the observed trajectory. Such changes may arise from parameter drift, environmental fluctuations, or transitions between different operating regimes, motivating Hamiltonian learning methods that can adapt to non-stationary quantum dynamics. 

The main contributions of this work are summarized as follows:
\begin{itemize}
    \item We formulate Hamiltonian learning as a QRL problem, where an agent infers the parameters of an unknown Hamiltonian from a sequence of time-evolved quantum states.
    \item We investigate non-stationary Hamiltonian learning under different forms of changing system dynamics, including a single abrupt switch, Hamiltonian changes at every time step, random abrupt switching over longer trajectories, and continuous gradual Hamiltonian drift.
    \item We investigate a recurrent QRL agent based on recurrent neural networks (RNNs) to assess the role of temporal memory in tracking changing Hamiltonian dynamics, and compare their performance with a non-recurrent QRL agent.
    \item We compare the proposed QRL approach with QGRNN, an existing quantum Hamiltonian learning method, to evaluate their performance under stationary and non-stationary dynamics.
    \item We evaluate the framework across different system sizes and dynamical regimes, and further assess the trained policies on previously unseen quantum-state trajectories without additional reward feedback, providing evidence that the learned policies generalize beyond the trajectories encountered during training.
\end{itemize}

The rest of the paper is organized as follows. Section~\ref{sec:Background} presents the necessary theoretical background. Section~\ref{sec:methodology} describes the proposed Hamiltonian learning approach, including the problem formulation, the QRL environment, and the considered agent architectures. Section~\ref{sec:Setup} presents the experimental setup, including the quantum environment, the evaluated Hamiltonian learning scenarios, and the agent hyperparameters. Section~\ref{sec:metrics} introduces the evaluation metrics used to assess the proposed approach. Section~\ref{sec:results} presents and discusses the experimental results. Section~\ref{sec:engineering} discusses engineering and implementation considerations, and finally Section~\ref{sec:conclusion} concludes the paper.

\section{Preliminaries}\label{sec:Background}

This section presents the theoretical foundations required for this work. We first introduce the classical Ising model, followed by the transverse-field Ising model and the fundamentals of QRL.

\subsection{Classical Ising Model}

The classical Ising model is widely used to model magnetic materials and study their phase transitions \cite{cipra1987introduction}. It represents a material as a lattice (graph) of particles, where each particle is associated with a spin variable that can take one of two values, $\sigma_{i,j}= \pm 1$ \cite{bian2010ising, brush1967history}. The spins interact with their neighboring particles, allowing the model to capture the magnetic properties of the material.

Here, we consider a two-dimensional square lattice in which each node interacts only with its four nearest neighbors. The Hamiltonian of the classical Ising model is given by \cite{cipra1987introduction,bian2010ising,choi2021tutorial}:

\begin{equation}
H = -J \sum_{(i,j)} \sigma_i \sigma_j - B\sum_i h_i \sigma_i,
\end{equation}
where $J$ denotes the interaction strength between neighboring spins, $B$ is the magnetic moment, and $h_i$ is the external magnetic field.

\subsection{Transverse-field Ising Model}\label{sec:TFIM}

The quantum counterpart of the classical Ising model is the transverse-field Ising model (TFIM). Unlike the classical Ising model, where each spin is restricted to one of two discrete values ($\pm1$), the quantum model represents each spin as a qubit that can exist in a superposition of states. TFIM captures effects that classical Ising models fail to explain, such as quantum phase transitions \cite{dutta2010quantum}. 

The TFIM Hamiltonian is defined as follows \cite{choi2021tutorial}:

\begin{equation}\label{TIMeq}
 H = J \sum_{(i, j )} \sigma_i^z \sigma_j^z - B \sum_i h_i \sigma_i^z - B\sum_i g_i \sigma_i^x
\end{equation}

\noindent where $\sigma_i^z$ and $\sigma_j^z$ are Pauli-Z operators acting on $i$ and $j$, respectively, and $\sigma_i^x$ is a Pauli-X operator acting on $i$.

The parameters $J$, $B$ and $h_i$ are as defined in the classical Ising model equation, and $g_i$ is the longitudinal magnetic field.

\subsection{Quantum Reinforcement Learning}

In reinforcement learning (RL), an agent learns by interacting with an environment through trial and error. At each discrete time step $t$, the agent observes the current state $s_t$, selects an action $a_t$, transitions to the next state $s_{t+1}$, and receives a scalar reward $r_t$. The agent's behavior is governed by a policy $\pi$, which is typically stochastic and maps each state to a probability distribution over the available actions:

\begin{equation}
\pi(a \mid s) = \Pr\left(a_t = a \mid s_t = s\right).
\end{equation}

The goal of RL is to learn a policy, parameterized by $\boldsymbol{\theta}$, that maximizes the expected return. RL algorithms are generally categorized into two groups: value-based and policy-based methods. Value-based methods learn a value function and derive a policy by selecting actions that maximize the estimated value. In contrast, policy-based methods directly optimize the policy parameters without learning an explicit value function.

QRL extends classical RL by incorporating quantum computation into one or more components of the learning process. Depending on the formulation, the environment, the agent, or both may be quantum. In a quantum environment, the agent interacts with quantum systems and receives quantum information, such as quantum states or measurement outcomes. A common QRL approach is to represent the policy using a parameterized quantum circuit. Such quantum policies are particularly well suited to quantum environments, where the observations available to the agent arise from quantum systems and may therefore retain intrinsically quantum characteristics \cite{broughton2020tensorflow}. In this work, we adopt both a quantum environment and a quantum agent because otherwise, when the agent is quantum, and the environment is classical, or provides classical observations, an inefficient encoding step is required to map the observations into quantum states before they can be processed by a parameterized quantum circuit. Instead, in our setting, the quantum environment directly provides the agent with time-evolved quantum states, while the agent employs a parameterized quantum circuit to learn the policy (or Hamiltonian parameters).


\section{Methodology} \label{sec:methodology}

This section presents the proposed Hamiltonian learning approach. We first formulate the Hamiltonian learning problem, then describe its formulation as a QRL environment, followed by the considered agent architectures and QGRNN baseline.

\subsection{Problem Formulation} \label{sec:problemFormulation}
Consider a closed quantum system governed by an unknown Hamiltonian $H$. Starting from an initial state $|\psi(0)\rangle$ at time step $t=0$, the system evolves to produce a sequence of quantum states (trajectory),

\[
|\psi(0)\rangle
\xrightarrow{\,H\,}
|\psi(1)\rangle
\xrightarrow{\,H\,}
|\psi(2)\rangle
\xrightarrow{\,H\,}
\dots
\xrightarrow{\,H\,}
|\psi(T)\rangle.
\]

where $T$ is the total number of time steps. The evolution of these states is governed by the Schrödinger equation. Given the state $|\psi (t)\rangle$ at time step $t$, the subsequent state is obtained as:

\begin{equation}
|\psi({t+1})\rangle
=
e^{-iH\Delta t}
|\psi (t)\rangle,
\end{equation}

where $\Delta t$ is the evolution time step.

In this work, the unknown Hamiltonian is modeled using the simplified TFIM, given by \cite{choi2021tutorial}:

\[
H(\boldsymbol{\alpha})
=
\sum_{(i, j )} {\alpha}^{(1)}_{(i, j )} \sigma_i^z \sigma_j^z +   \sum_i  {\alpha}^{(2)}_{(i)} \sigma_i^z +\sum_i  \sigma_i^x 
\]

where $\boldsymbol{\alpha}=\{\alpha^{(1)},\alpha^{(2)}\}$ denotes the unknown Hamiltonian parameters. For an $n$-qubit system, the Hamiltonian contains $2n$ unknown parameters. Therefore, the dimensionality of the parameter estimation problem grows with the system size.

The Hamiltonian is assumed to be inaccessible to the agent. Instead, the agent receives only the sequence of time-evolved states and must infer the underlying Hamiltonian parameters solely from these states.

While the above formulation assumes a stationary Hamiltonian, many practical quantum systems exhibit varying dynamics due to parameter drift, environmental fluctuations, or changes in external control fields. In such settings, the Hamiltonian governing the system may change during the observed trajectory, thus affecting how the states evolve. These changes may occur abruptly or gradually over time. To model such scenarios, we consider multiple Hamiltonians, denoted by $H_1, H_2, \ldots$, where each Hamiltonian is defined by a different set of parameters. For example, a system may evolve under Hamiltonian $H_1$ during the first portion of the trajectory before switching abruptly to Hamiltonian $H_2$ at a switch time $t_s$, 
\begin{equation} \label{eq:switching}
H(t)=
\begin{cases}
H_1, & t<t_s,\\
H_2, & t\ge t_s,
\end{cases}
\end{equation}

where $H_1$ and $H_2$ denote Hamiltonians characterized by different parameter sets. Suppose the transition occurs at $t_s = 2$, the system evolves according to:

\[
|\psi(0)\rangle
\xrightarrow{\,H_1\,}
|\psi(1)\rangle
\xrightarrow{\,H_1\,}
|\psi(2)\rangle
\xrightarrow{\,H_2\,}
\dots
\xrightarrow{\,H_2\,}
|\psi(T)\rangle.
\]


More generally, the governing Hamiltonian may undergo multiple abrupt changes or vary gradually throughout a single trajectory. In this paper, we investigate several stationary and non-stationary scenarios to assess the ability of the proposed QRL approach to identify Hamiltonian parameters under different forms of system dynamics.

\subsection{Hamiltonian Learning as a QRL Environment}

\begin{figure*}[!t]
    \centering
    \includegraphics[width=1\linewidth]{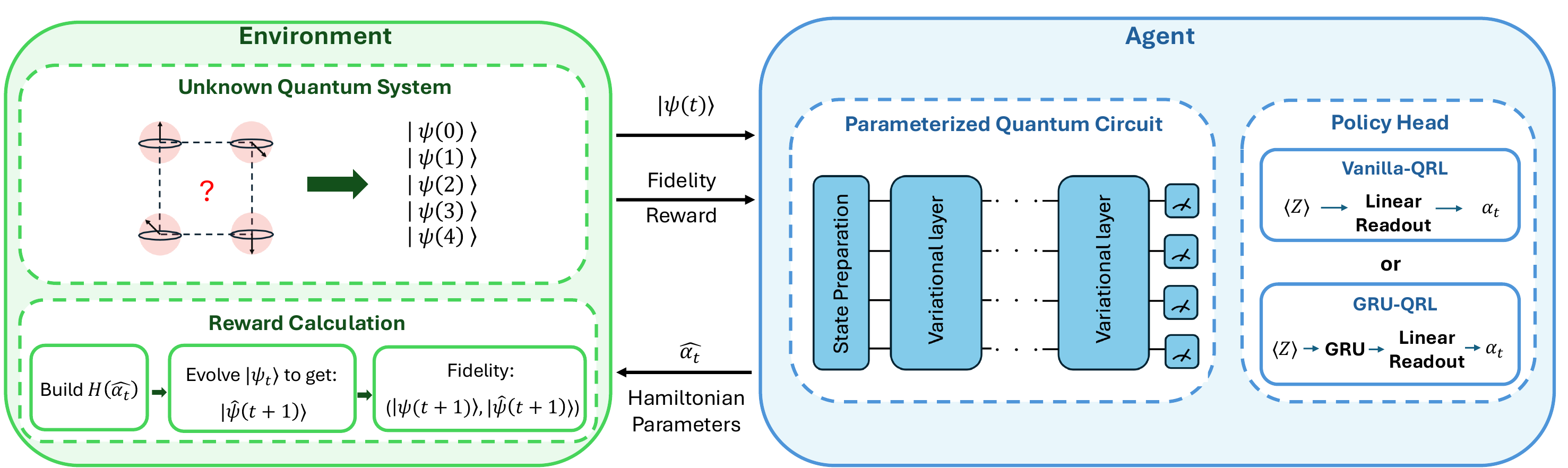}
    \caption{Overview of the proposed quantum reinforcement learning framework for Hamiltonian learning. The environment provides sequential quantum states, while the agent estimates the Hamiltonian parameters using either the Vanilla-QRL or GRU-QRL policy. The reward is computed from the fidelity between the predicted and target next quantum states.}
    \label{fig:interaction}
\end{figure*}

To infer the unknown Hamiltonian parameters, we formulate Hamiltonian learning as a QRL problem, in which an agent interacts with a quantum environment through a sequence of episodes. During each episode, the agent sequentially observes time-evolved quantum states trajectory generated by the unknown system and produces estimates of the Hamiltonian parameters. Based on these estimates, the environment evaluates the predicted system Hamiltonian and returns a reward that guides the learning process. The interaction continues until the episode terminates, enabling the agent to progressively improve its parameter estimation policy through repeated interactions with the environment. Figure~\ref{fig:interaction} illustrates the interaction between the agent and the environment. The episode structure and the interaction performed at each time step are defined as follows.

\paragraph{Episode}
An episode consists of one complete interaction between the agent and the quantum environment. During an episode, the environment is provided access to a trajectory of time-evolved quantum states generated by the unknown quantum system. In practice, these states are assumed to be collected from the physical quantum system; however, in this work, they are simulated. The trajectory is given by

\begin{equation}
\left\{
|\psi(0)\rangle,
|\psi(1)\rangle,
\ldots,
|\psi(T)\rangle
\right\}.
\end{equation}

The agent processes this trajectory sequentially, receiving one quantum state at each time step $t$.

\paragraph{Time Step}

At each time step $t$, the interaction between the agent and the environment proceeds as follows:

\begin{enumerate}
    \item The environment provides the agent with the current time-evolved quantum state,
    \[
    o_t = |\psi(t)\rangle,
    \]
    which serves as the observation.

    \item Based on the observation, the agent selects an action,
    \[
    a_t = \hat{\boldsymbol{\alpha}}_t,
    \]
    where the action corresponds to the estimated Hamiltonian parameters. 
    
    \item The environment then uses these parameters to construct the estimated Hamiltonian $\hat{H}$.

    \item Using the estimated Hamiltonian $\hat{H}$, the environment evolves the current state $|\psi(t)\rangle$ to obtain the predicted next state,
    \[
    |\hat{\psi}(t+1)\rangle.
    \]

    \item The environment computes the reward by comparing the predicted next state $|\hat{\psi}(t+1)\rangle$ with the true next state $|\psi(t+1)\rangle$ using the state fidelity.

    \item The environment advances to the next state in the trajectory, and the process repeats until the final state has been processed, and thus completing a full episode.
\end{enumerate}

The reward at each time step is defined as the quantum state fidelity between the predicted next state $|\hat{\psi}(t+1)\rangle$ and the true next state $|\psi(t+1)\rangle$,

\begin{equation} \label{eq:reward}
r_t = F\!\left(|\psi(t+1)\rangle, |\hat{\psi}(t+1)\rangle\right),
\end{equation}

where the state fidelity satisfies

\begin{equation}
0 \leq F \leq 1.
\end{equation}

A fidelity of $F=1$ indicates that the predicted and true quantum states are identical, whereas $F=0$ indicates that they are orthogonal. The fidelity is adopted as the reward because the objective of Hamiltonian learning is to estimate Hamiltonian parameters that accurately reproduce the evolution of the unknown quantum system. Maximizing the fidelity encourages the agent to select actions that generate predicted states closely matching the true system dynamics. In other words, actions producing inaccurate Hamiltonian estimates receive lower rewards, guiding the policy toward parameter estimates that better capture the underlying Hamiltonian.

\subsection{Agent Architecture}
We first develop a simple QRL agent, referred to as \textit{Vanilla-QRL}. While this architecture is suitable for stationary Hamiltonian learning, extending it to non-stationary dynamics presents additional challenges, as the agent has no explicit mechanism for retaining information from previously observed quantum states. To address this limitation, we propose a second variant, called \textit{GRU-QRL}, which augments the policy network with a Gated Recurrent Unit (GRU). By maintaining an internal memory of past observations, GRU-QRL is designed to capture temporal dependencies in the quantum state trajectory, enabling more effective learning under changing Hamiltonian dynamics.

\begin{figure}
    \centering
    \includegraphics[width=1\linewidth]{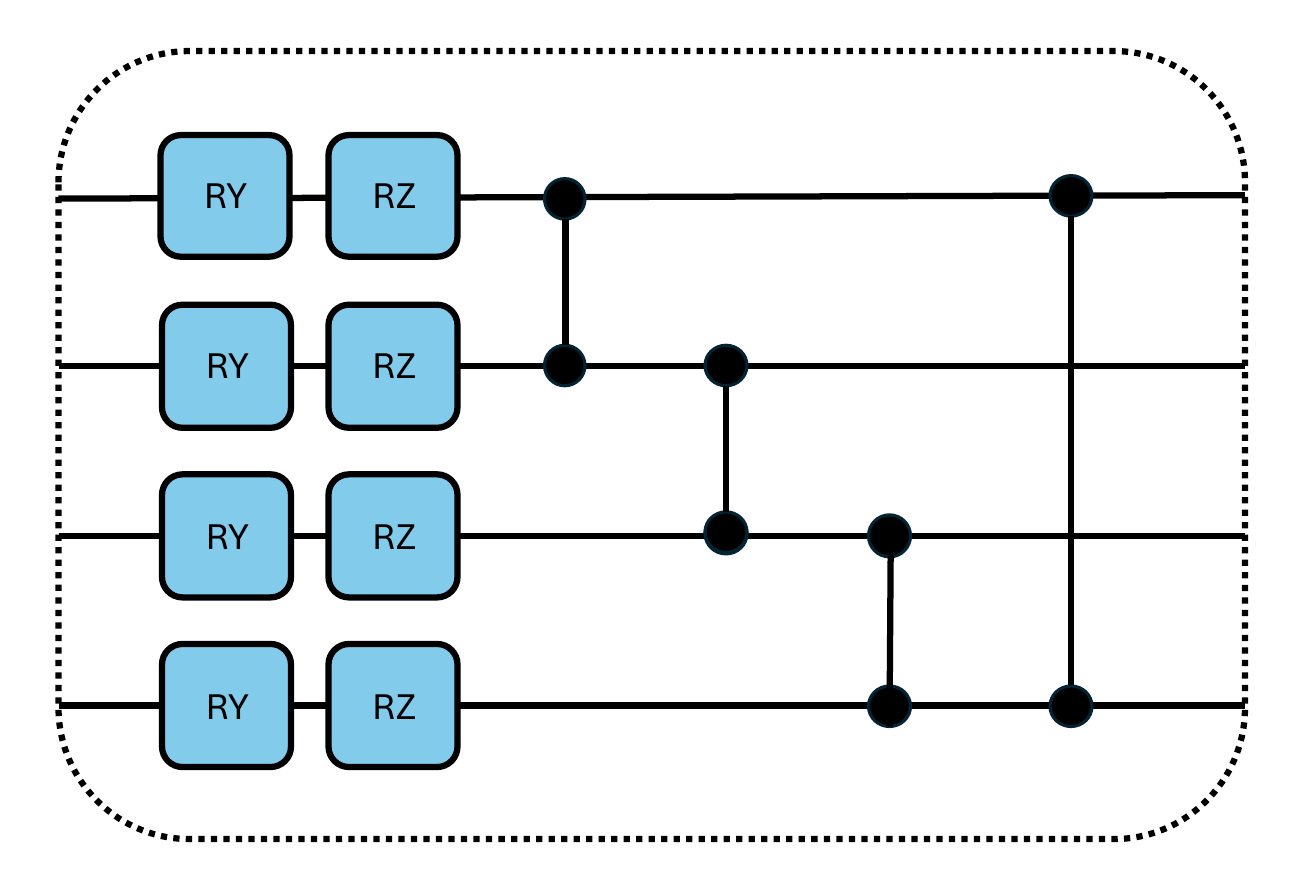}
    \caption{A single layer of the parameterized quantum circuit used by both policies, illustrated for the 4-qubit case. The number of qubits scales with the system size.}
    \label{fig:variational_block}
\end{figure}

Although the two agent variants differ in their policy heads, they share the same parameterized quantum circuit. As illustrated in Figure~\ref{fig:interaction}, each time-evolved quantum state is fed into the circuit using a state preparation layer. The prepared state is then processed by a variational block, shown in Figure~\ref{fig:variational_block}, consisting of parameterized $R_Y$ and $R_Z$ rotation gates applied to each qubit, followed by a ring of controlled-$Z$ (CZ) gates. The number of qubits in the circuit is chosen to match the number of qubits in the target quantum system. The variational block is repeated according to the circuit depth, $L$. Finally, Pauli-$Z$ expectation values are measured from all qubits and passed to the policy head, where the two agent variants differ.

\subsubsection{Vanilla-QRL}

The Vanilla-QRL agent maps each observed quantum state directly to an estimate of the Hamiltonian parameters, without retaining any information from previous time steps. 

The quantum states are loaded into the circuit, and after applying $L$ variational layers, the circuit returns the expectation value of the Pauli-Z operator on each qubit, $\langle Z\rangle _t =\left[\langle Z_1\rangle,\langle Z_2\rangle,\ldots,\langle Z_n\rangle\right]$, yielding an $n$-dimensional classical measurement vector. The measurement vector is passed through a linear layer mapping the $n$ measurements to the $2n$-dimensional action space (matching the $2n$ Hamiltonian parameters under the TFIM formulation). The policy is trained via REINFORCE using the fidelity-based reward defined in eq.~\ref{eq:reward}.

\subsubsection{GRU-QRL}

GRU-QRL introduces a recurrent memory between the quantum measurement and the classical readout layer, allowing the policy to condition its estimate at time $t$ on information accumulated from all previously observed states in the trajectory. At each time step $t$, the $n$-dimensional measurement vector  $\langle Z\rangle _t=\left[\langle Z_1\rangle,\langle Z_2\rangle,\ldots,\langle Z_n\rangle\right]$ produced by the parameterized quantum circuit is passed, together with the hidden state $h_{t-1}$ from the previous time step, into a GRU cell:

\begin{equation}
h_t = \mathrm{GRU}\!\left(\langle Z \rangle_t,\, h_{t-1}\right),
\end{equation}

where $h_t$ is the updated hidden state. The hidden state $h_t$ rather than the raw measurement vector, is passed through the same linear readout used in Vanilla-QRL, producing the action. 

As in Vanilla-QRL, the policy is trained via REINFORCE using the fidelity-based reward.

\subsection{QGRNN Baseline}\label{sec:qgrnn}

To provide a quantum Hamiltonian learning baseline, we compare the proposed QRL approach with the QGRNN introduced in \cite{verdon2019quantum}. QGRNN was introduced to learn the Hamiltonian of an unknown quantum system from its time-evolved quantum states. Therefore, QGRNN is closely aligned with our approach in terms of the available data, the Hamiltonian model and the overall objective, making it a natural baseline for comparison. Unlike the proposed QRL formulation, QGRNN directly parameterizes the Hamiltonian governing the predicted quantum evolution, where the trainable parameters correspond to the interaction strengths and local fields of the candidate Hamiltonian. 

QGRNN uses two quantum registers to compare the true and predicted quantum state evolution. For each transition, both registers are initialized with the current state $|\psi(t)\rangle$. The first register is evolved to the true next state $|\psi(t+1)\rangle$, while the second register is evolved using the parameterized QGRNN layers to produce the predicted next state $|\hat{\psi}(t+1)\rangle$. In this work, the true state evolution is simulated, while in practice, these states would be obtained from the physical quantum system. The two resulting states are then compared using a SWAP test. The resulting fidelity is used to define the loss and update the QGRNN parameters through gradient descent.

QGRNN is trained using the same time-evolved quantum state trajectories considered in this work. The state transitions are processed in temporal order, where the QGRNN parameters are updated through gradient descent after each transition. Under non-stationary dynamics, the parameters are therefore continuously updated as the governing Hamiltonian changes throughout the trajectory.

\section{Experimental Setup} \label{sec:Setup}

This section describes the experimental setup used throughout this work, including the quantum environment, the evaluated Hamiltonian learning scenarios, the QRL agents with their corresponding hyperparameters, and the QGRNN hyperparameters.

\subsection{Environment}

We evaluate the proposed framework on TFIM of $n \in \{2,4,7\}$ qubits, with the target Hamiltonian defined over a closed loop topology. For each system size, we generate $300$ independent trajectories, each initialized from a randomly sampled initial state $|\psi(0)\rangle$. The evolution time step between consecutive states is fixed at $\Delta t=0.25$.

The target Hamiltonian parameters $\boldsymbol{\alpha}$ used to generate each trajectory are fixed per system size across all $300$ trajectories for a given scenario. This setting reflects a practical Hamiltonian learning task, where multiple experiments are performed on the same physical quantum system. Although each experiment begins from a different initial state, the underlying Hamiltonian remains unchanged, allowing the agent to observe diverse system dynamics while learning the same unknown Hamiltonian. For the non-stationary scenarios, the Hamiltonians defining the changing dynamics are likewise fixed per system size, while the initial states and, where applicable, the switching behavior vary across trajectories (scenario 4).

In this work, we consider five experimental scenarios, distinguished by how the governing Hamiltonian evolves over the course of a trajectory. Scenarios 1-3 use trajectories of $T=4$ time steps, corresponding to four state transitions and five quantum states, while Scenarios 4 and 5 consider longer trajectories of $T=20$ time steps, corresponding to 20 state transitions and 21 quantum states.

\paragraph{Scenario 1: Stationary Hamiltonian}The system evolves under a single, fixed Hamiltonian $H$ throughout the entire trajectory:
\[
|\psi(0)\rangle
\xrightarrow{\,H\,}
|\psi(1)\rangle
\xrightarrow{\,H\,}
|\psi(2)\rangle
\xrightarrow{\,H\,}
|\psi(3)\rangle
\xrightarrow{\,H\,}
|\psi(4)\rangle.
\]

\paragraph{Scenario 2: Single Abrupt Switch}The system evolves under $H_1$	for the first portion of the trajectory, then switches abruptly to $H_2$ at a fixed switch step $t_s$, following the piecewise formulation in Eq.~\ref{eq:switching}. In our experiments, the switch occurs at the midpoint of the trajectory, $t_s=2$, so that the agent observes two transitions generated under $H_1$ followed by two transitions generated under $H_2$:
\[
|\psi(0)\rangle
\xrightarrow{\,H_1\,}
|\psi(1)\rangle
\xrightarrow{\,H_1\,}
|\psi(2)\rangle
\xrightarrow{\,H_2\,}
|\psi(3)\rangle
\xrightarrow{\,H_2\,}
|\psi(4)\rangle.
\]

This scenario evaluates the agent's ability to adapt to an unannounced change in system dynamics and adapt its parameter estimate accordingly, without any explicit signal indicating that a switch has occurred.

\paragraph{Scenario 3: Every-Step Switching} The system evolves under a distinct Hamiltonian at every time step:

\[
|\psi(0)\rangle
\xrightarrow{\,H_1\,}
|\psi(1)\rangle
\xrightarrow{\,H_2\,}
|\psi(2)\rangle
\xrightarrow{\,H_3\,}
|\psi(3)\rangle
\xrightarrow{\,H_4\,}
|\psi(4)\rangle.
\]

where $H_1$, $H_2$, $H_3$ and $H_4$ denote four distinct Hamiltonians, each characterized by its own set of parameters $\boldsymbol{\alpha}_1$, $\boldsymbol{\alpha}_2$, $\boldsymbol{\alpha}_3$ and $\boldsymbol{\alpha}_4$. This scenario represents a highly non-stationary setting, as the agent must continuously re-estimate the governing Hamiltonian at every step, with no interval of stationary dynamics on which to stabilize its estimate.

\paragraph{Scenario 4: Longer Trajectory with Random Abrupt Switching}

Scenario 4 extends the single abrupt switching setting introduced in scenario 2 to a longer trajectory of $T=20$ time steps. The system still evolves under $H_1$ before abruptly switching to $H_2$, following the formulation in Eq. \ref{eq:switching}. However, unlike scenario 2, where switching occurs at a fixed midpoint, the switching location varies across trajectories, making it more challenging.  

The Hamiltonian switch is allowed to occur randomly at $t_s \in \{5,10,15\}$. For each trajectory, one of these switching locations is randomly selected, resulting in trajectories where the system may switch relatively early, around the middle, or later in the sequence. The switch location is not provided to the agent, and the GRU-QRL agent must therefore adapt to the change in the system dynamics from the observed sequence of quantum states. 

\paragraph{Scenario 5: Longer Trajectory with Continuous Gradual Hamiltonian Drift}

In the previous non-stationary scenarios, the transition between Hamiltonians occurs abruptly at a particular time step. In scenario 5, we instead consider a gradual change in system dynamics over a longer trajectory of $T=20$ time steps. The system begins under Hamiltonian $H_1$, and its parameters gradually transition towards those of $H_2$ throughout the trajectory. 

The Hamiltonian governing the system at each time step is calculated as:

\begin{equation}
\begin{aligned}
H(t) &= (1-\lambda_t)H_1 + \lambda_t H_2, \\
\end{aligned}
\end{equation}

where $\lambda_t = \frac{t}{T-1}$ and $\quad t = 0,\ldots,T-1$. For $T=20$, the state trajectory evolves under a continuously changing Hamiltonian:

\[
|\psi(0)\rangle
\xrightarrow{H(0)}
|\psi(1)\rangle
\xrightarrow{H(1)}
|\psi(2)\rangle
\cdots
\xrightarrow{H(19)}
|\psi(20)\rangle.
\]

$\lambda_t$ gradually increases from $0$ to $1$ across the trajectory. Therefore, the Hamiltonian changes slightly at every time step, progressively moving from $H_1$ towards $H_2$. At the beginning, $\lambda_0=0$ and hence $H(0)=H_1$, while at the final transition, $\lambda_{19}=1$ and $H(19)=H_2$. Thus, unlike the abrupt switching scenarios, there is no single switching point; instead, the system dynamics continuously drift throughout the trajectory.

We evaluate all scenarios across system sizes $n \in \{2,4,7\}$, allowing us to assess how the agent's Hamiltonian parameter estimation accuracy is affected by the size of the underlying quantum system.

\subsection{Agents}

The proposed framework is evaluated using two QRL agent variants: Vanilla-QRL and GRU-QRL. We first develop and evaluate the Vanilla-QRL agent on the stationary Hamiltonian learning task (Scenario 1). The Vanilla-QRL agent is then further evaluated under the single abrupt switching scenario (Scenario 2) to assess its ability to operate under non-stationary dynamics. Motivated by the limitations of the feedforward policy, we introduce GRU-QRL. 
The hyperparameters used for both agents are summarized in Table~\ref{tab:hyperparameters}. These hyperparameters were selected empirically based on preliminary experiments to provide stable training performance across the considered scenarios.

\begin{table}[t]
\centering
\caption{Experimental hyperparameters for all evaluated agents}
\label{tab:hyperparameters}
\begin{tabular}{lcc}
\hline
\textbf{Hyperparameter} & \textbf{Vanilla-QRL} & \textbf{GRU-QRL} \\
\hline
\hline
Depth ($L$)              & 3      & 3 \\
Number of Circuit Qubits         & $n$    & $n$ \\
Learning Rate                & 0.01  & 0.01 \\
Optimizer                    & Adam   & Adam \\
Training Epochs              & 300    & 300 \\
Episodes per Epoch                & 20     & 20 \\
GRU Hidden Dimension         & --     & 16 \\

\hline
\end{tabular}
\end{table}

\subsection{QGRNN Baseline Setup}

QGRNN is evaluated using the same system sizes, Hamiltonian parameters, initial-state generation procedure, and quantum state trajectories used for the QRL experiments.  The hyperparameters used for QGRNN is summarized in Table~\ref{tab:qgrnn_hyperparameters}.

\begin{table}[t]
\centering
\caption{QGRNN (baseline) hyperparameters}
\label{tab:qgrnn_hyperparameters}
\begin{tabular}{lc}
\hline
\textbf{Hyperparameter} & \textbf{Value} \\
\hline
\hline
Depth & 25 \\
Optimizer & Adam \\
Learning Rate & 0.5 \\
Training Epochs & 300 \\
Trotter Step & 0.01 \\
\hline
\end{tabular}
\end{table}

It is important to note that the QGRNN depth was not selected arbitrarily. Following the QGRNN construction, the number of layers is determined by the evolution time and the Trotter step size. Since the evolution time between consecutive states is $\Delta t=0.25$ and the Trotter step is set to $0.01$, the resulting circuit depth is

\begin{equation}
L = \frac{\Delta t}{\text{Trotter Step}}
= \frac{0.25}{0.01}
= 25.
\end{equation}

Therefore, the depth of 25 reported in Table~\ref{tab:qgrnn_hyperparameters} follows directly from the QGRNN evolution construction rather than being treated as a separately tuned hyperparameter.

The learning rate of $0.5$ was adopted from the established QGRNN literature, following \cite{kaldari2026feature} and \cite{Ceroni_2020}.

\section{Evaluation Metrics} \label{sec:metrics}

To quantitatively evaluate the proposed framework, we consider both the accuracy of the reconstructed quantum dynamics and the accuracy of the recovered Hamiltonian parameters. Specifically, we report the following metrics.

\paragraph{Fidelity}

To evaluate the accuracy of the learned Hamiltonian dynamics, we compute the quantum state fidelity between the target evolved state $|\psi_t\rangle$ and the state $|\hat{\psi}_t\rangle$ generated using the estimated Hamiltonian. The final fidelity is obtained by averaging the fidelity over the final evaluation trajectories. A fidelity of one indicates perfect agreement between the target and reconstructed quantum states.

\paragraph{Parameter estimation error} next is the mean squared error (MSE) between the true and estimated Hamiltonian parameters. Let $\alpha_{\text{target}} \in \mathbb{R}^d$ denote the true Hamiltonian parameter vector and $\hat{\alpha} \in \mathbb{R}^d$ denote the corresponding estimated parameter vector produced by the agent, where $d$ is the total number of Hamiltonian parameters. The parameter estimation MSE is defined as
\begin{equation}
    \text{MSE} = \frac{1}{d}\sum_{k=1}^{d} \left(\alpha_{\text{target},k} - \hat{\alpha}_k\right)^2.
\end{equation}
For systems with $n \geq 3$ qubits, $d = 2n$, consistent with the parameterization introduced in Section \ref{sec:problemFormulation}. For the $n=2$ case, the cycle graph reduces to a single interaction edge rather than two, so $d = 3$.

For the non-stationary scenarios, where the system evolves under multiple Hamiltonians $H_1, H_2, \dots$, the agent's estimated parameters are averaged separately over the time steps corresponding to each Hamiltonian segment, yielding segment-wise estimates $\hat{\alpha}_1, \hat{\alpha}_2, \dots$ In this setting, we report the combined MSE computed over the concatenation of all segments,
\begin{equation}
    \text{MSE}_{\text{combined}} = \frac{1}{Md}\sum_{k=1}^{Md} \left(\alpha_{\text{target},k} - \hat{\alpha}_k\right)^2,
\end{equation}

where $\alpha_{\text{target}} = \left[\alpha_1^{\text{target}} \,\|\, \alpha_2^{\text{target}} \,\|\, \dots \right]$ and $\hat{\alpha} = \left[\hat{\alpha}_1 \,\|\, \hat{\alpha}_2 \,\|\, \dots\right]$ denote the concatenated target and estimated parameter vectors across all $M$ Hamiltonian segments in the trajectory (for example, $M=2$ for Scenario 2, $M=4$ for Scenario 3).

For the gradual Hamiltonian drift scenario, we calculate the parameter estimation error across all time steps. The drift MSE is defined as:

\begin{equation}
    \text{MSE}_{\text{drift}} =
    \frac{1}{Td}
    \sum_{t=0}^{T-1}
    \sum_{k=1}^{d}
    \left(\alpha_k(t)-\hat{\alpha}_k(t)\right)^2,
\end{equation}

where $T$ is the total number of time steps, $\alpha_k(t)$ denotes the target value at time step $t$, and $\hat{\alpha}_k(t)$ denotes the corresponding estimated parameter. 

\section{Results} \label{sec:results}

Our goal is to evaluate the proposed QRL framework under both stationary and non-stationary Hamiltonian learning scenarios as the quantum system scales.

We begin by evaluating the Vanilla-QRL agent on the stationary Hamiltonian learning task. As the learning scenarios become increasingly challenging, we compare the performance of Vanilla-QRL and GRU-QRL, and subsequently adopt the GRU-QRL agent for the most challenging scenarios due to its superior empirical performance. Throughout, we compare against the QGRNN baseline to assess how each approach performs relative to an existing quantum Hamiltonian learning method. 

All reported results are averaged over three independent random seeds, with the mean and standard deviation reported where applicable.

\subsection{Scenario 1: Stationary Hamiltonian}

We first evaluate the proposed Vanilla-QRL agent on the stationary Hamiltonian learning task, where the underlying Hamiltonian remains fixed throughout the entire trajectory. This serves as the baseline scenario for assessing the agent's ability to accurately recover the target Hamiltonian before considering more challenging non-stationary settings.

\begin{figure}[h]
    \centering
    \includegraphics[width=1\linewidth]{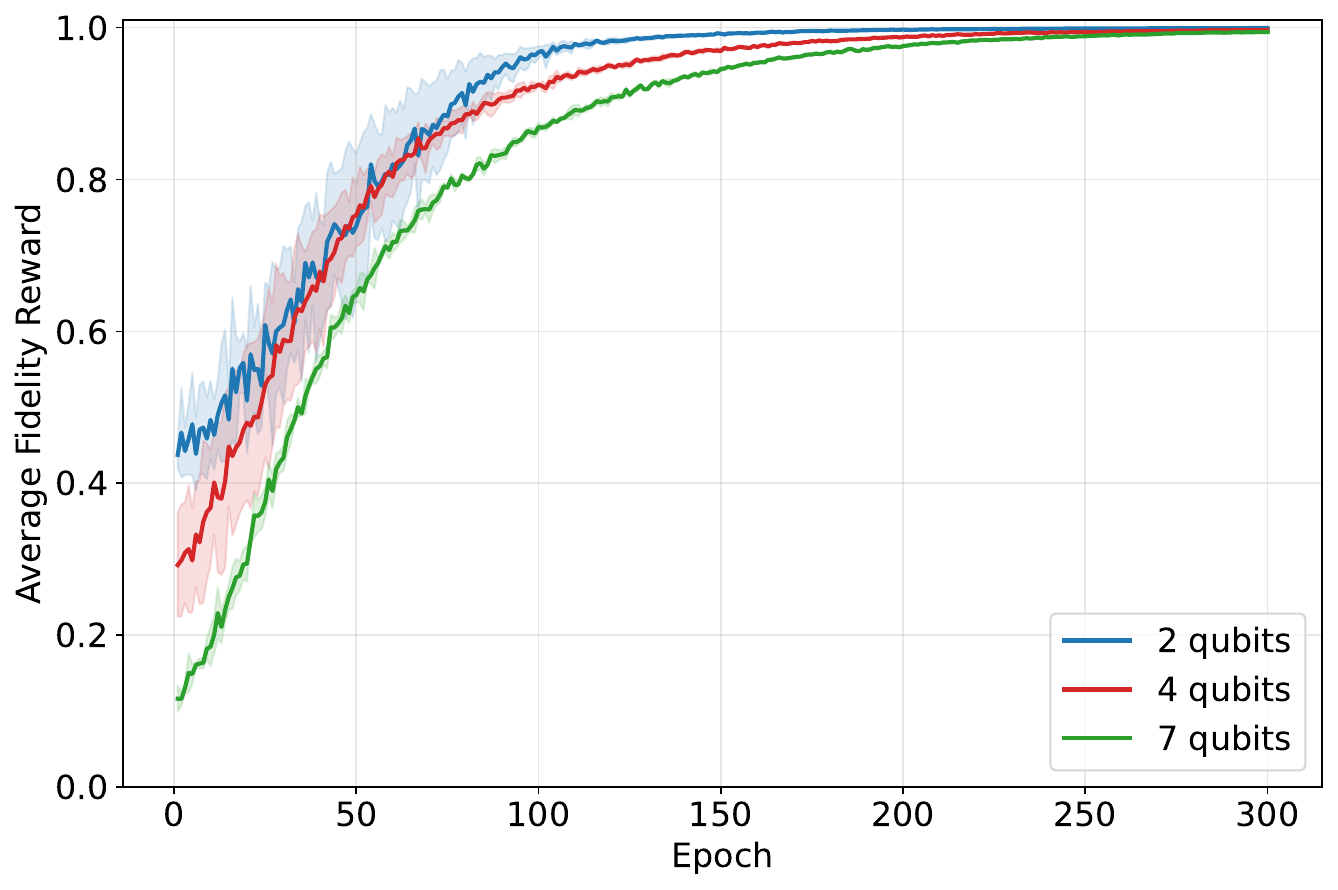}
    \caption{Learning curves of the Vanilla-QRL agent under Scenario 1 (Stationary Hamiltonian Learning). Shaded regions indicate the standard deviation over three independent random seeds.}
    \label{fig:Scenario(1)_curve}
\end{figure}

\begin{figure}[htbp]
    \centering

    \subfigure[2-qubit system]{
        \includegraphics[width=1\linewidth]{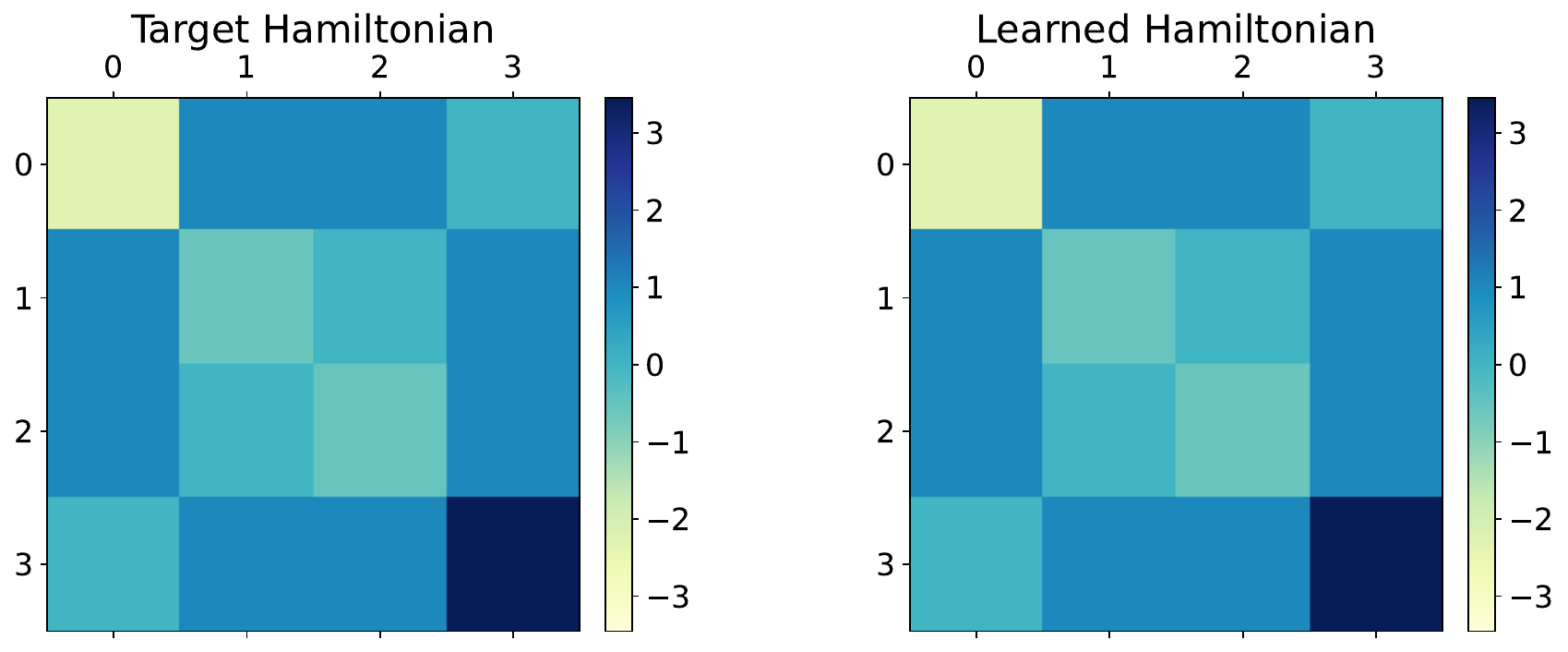}
        \label{fig:heatmap2}
    }
    \hfill
    \subfigure[4-qubit system]{
        \includegraphics[width=1\linewidth]{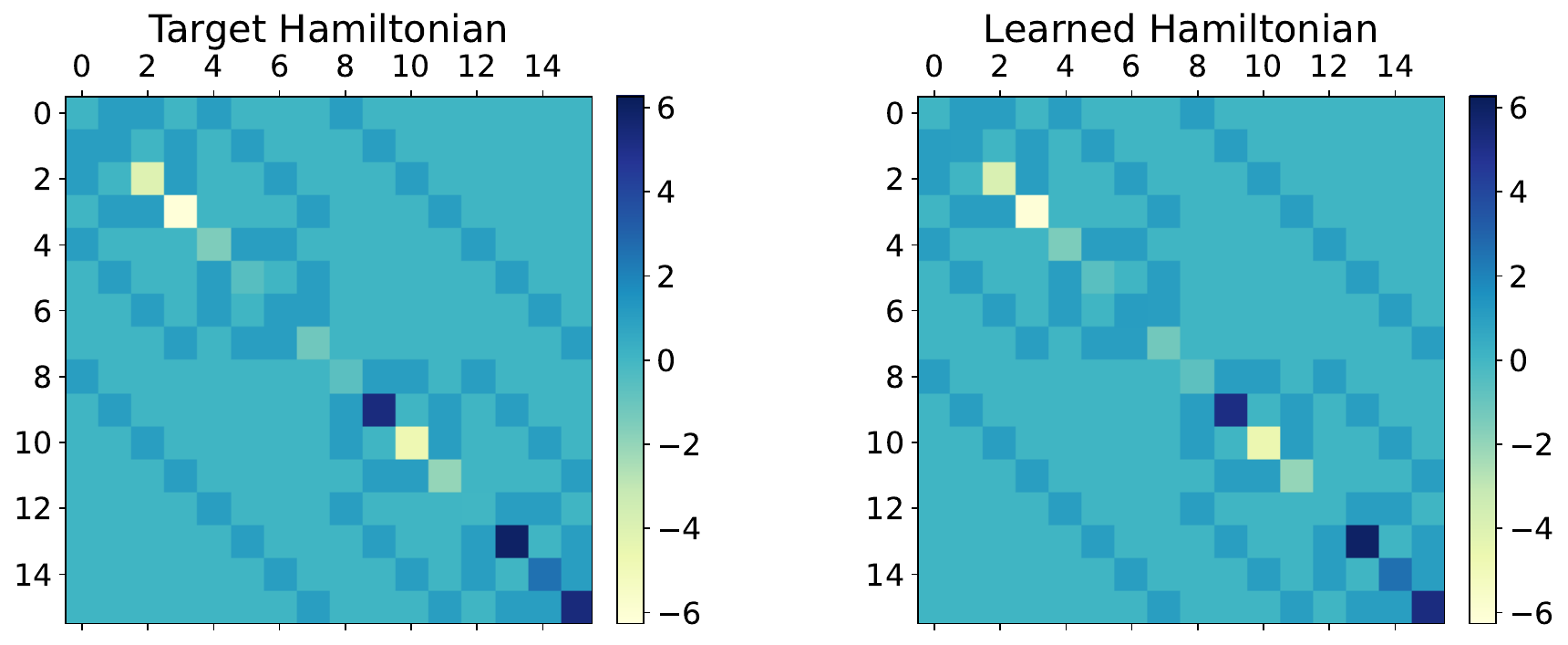}
        \label{fig:heatmap4}
    }

    \caption{Representative target and learned Hamiltonian matrix heatmaps obtained using Vanilla-QRL under Scenario 1 (Stationary Hamiltonian Learning).}
    \label{fig:heatmaps_stationary}
\end{figure}

Fig.~\ref{fig:Scenario(1)_curve} shows the training performance of Vanilla-QRL for systems with $n \in {2,4,7}$ qubits. In all cases, the fidelity reward increases steadily throughout training and converges to values close to one, demonstrating stable learning of the Hamiltonian parameters. These results indicate that the agent successfully reconstructs the Hamiltonian governing the observed quantum states dynamics. Although convergence becomes slightly slower as the system size increases, the agent consistently learns an accurate representation of the target Hamiltonian, with the final fidelity remaining close to one across all system sizes.

Fig.~\ref{fig:heatmaps_stationary} compares representative target and learned Hamiltonian matrix heatmaps for the 2- and 4-qubit systems. In both cases, the reconstructed Hamiltonians accurately preserve the structure of the target matrix elements, indicating that the learned parameters successfully recover the underlying system Hamiltonian. For the 7-qubit system, visual inspection through heatmaps becomes less informative due to the large Hamiltonian dimension ($128 \times 128$). Therefore, we rely on the quantitative MSE results reported in Table~\ref{tab:stationary_results}, which confirm accurate Hamiltonian reconstruction for the target system.

Table~\ref{tab:stationary_results} summarizes the quantitative performance of the proposed framework across all system sizes. The proposed QRL framework consistently achieves high final fidelity, exceeding $0.994$ in all cases, indicating that the learned Hamiltonians accurately reproduce the target quantum dynamics. Furthermore, the Hamiltonian reconstruction error remains low, with MSE values on the order of $10^{-3}$ or smaller across all system sizes. Although the system size increases from 2 to 7 qubits, resulting in a substantially larger Hamiltonian representation, only a slight reduction in fidelity is observed while maintaining a low reconstruction error. The small standard deviations across the three independent random seeds further demonstrate the robustness and stability of the proposed learning framework.

As a baseline comparison, QGRNN also achieves strong performance in stationary dynamics. As shown in Table~\ref{tab:stationary_results}, both Vanilla-QRL and QGRNN achieve fidelities above $0.994$ across all system sizes, while maintaining low parameter MSEs. These results show that both approaches are able to accurately learn the stationary Hamiltonian from the observed quantum state trajectories.

\begin{table*}[t]
\centering
\caption{Performance comparison of Vanilla-QRL and QGRNN under Scenario 1 (Stationary Hamiltonian Learning). Results are reported as mean $\pm$ standard deviation over three independent random seeds.}
\label{tab:stationary_results}
\setlength{\tabcolsep}{3pt}
\begin{tabular}{ccccc}
\hline
& \multicolumn{2}{c}{\textbf{Vanilla-QRL}}
& \multicolumn{2}{c}{\textbf{QGRNN}} \\
\cline{2-5}
\textbf{System Size}
& \textbf{Fidelity} & \textbf{MSE}
& \textbf{Fidelity} & \textbf{MSE} \\
\hline
2 qubits
& $0.999512 \pm 0.000098$
& $0.000452 \pm 0.000188$
& $ 0.999702 \pm 0.000241$
& $ 0.002009 \pm 0.001974 $ \\

4 qubits
& $0.997134 \pm 0.000499$
& $0.001288 \pm 0.000799$
& $0.996221 \pm 0.002302 $
& $0.002037 \pm 0.000393  $ \\

7 qubits
& $0.994119 \pm 0.000593$
& $0.001222 \pm 0.000336$
& $ 0.998406 \pm 0.000583 $
& $ 0.000108 \pm  0.000034$ \\
\hline
\end{tabular}
\end{table*}

\subsection{Scenario 2: Single Abrupt Hamiltonian Switch}

We next evaluate the proposed framework under the single abrupt switching scenario, where the system evolves under Hamiltonian $H_1$ during the first half of the trajectory before switching to a different Hamiltonian $H_2$ at the midpoint. Here, the governing system dynamics change without any explicit indication to the agent, requiring it to identify both Hamiltonians solely from the observed sequence of quantum states.

\begin{table*}[t]
\centering
\caption{Performance comparison of Vanilla-QRL, GRU-QRL and QGRNN under Scenario 2 (Single Abrupt Hamiltonian Switching). Results are reported as mean $\pm$ standard deviation over three independent random seeds.}

\label{tab:scenario2}
\setlength{\tabcolsep}{3pt}
\begin{tabular}{ccccccc}
\hline
& \multicolumn{2}{c}{\textbf{Vanilla-QRL}}
& \multicolumn{2}{c}{\textbf{GRU-QRL}}
& \multicolumn{2}{c}{\textbf{QGRNN}} \\
\cline{2-7}
\textbf{System Size}
& \textbf{Fidelity} & \textbf{MSE}
& \textbf{Fidelity} & \textbf{MSE}
& \textbf{Fidelity} & \textbf{MSE} \\
\hline

2 qubits
& $0.891023 \pm 0.008844$
& $0.739884 \pm 0.051994$
& $0.998263 \pm 0.000145$
& $0.003982 \pm 0.000280$
& $0.896083 \pm 0.033953$
& $0.785113 \pm 0.112755$ \\

4 qubits
& $0.828762 \pm 0.005340$
& $0.366573 \pm 0.015899$
& $0.989783 \pm 0.003433$
& $0.011661 \pm 0.006886$
& $0.772024 \pm 0.038214$
& $0.418885 \pm 0.071559$ \\

7 qubits
& $0.801367 \pm 0.002795$
& $0.213510 \pm 0.002079$
& $0.987590 \pm 0.000731$
& $0.005781 \pm 0.000728$
& $0.797496 \pm 0.015007$
& $0.235578 \pm 0.007727$ \\

\hline
\end{tabular}
\end{table*}

\subsubsection{Vanilla-QRL}

\begin{figure}[h]
    \centering
    \includegraphics[width=1\linewidth]{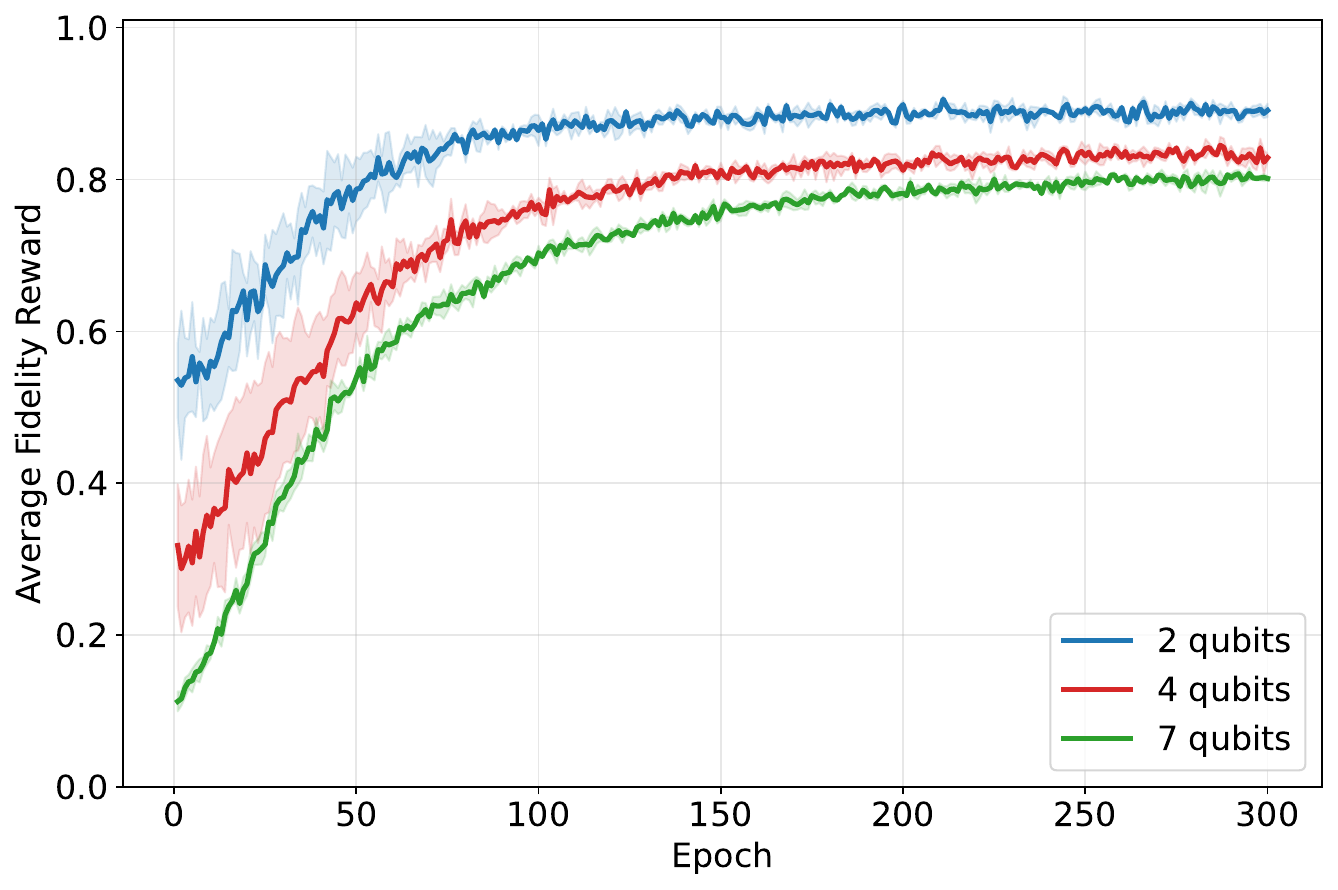}
    \caption{Learning curves of the Vanilla-QRL agent under Scenario 2 (Single Abrupt Hamiltonian Switching). Shaded regions indicate the standard deviation over three independent random seeds.}
    \label{fig:Scenario(2)_vanilla_curve}
\end{figure}

Fig.~\ref{fig:Scenario(2)_vanilla_curve} presents the training performance of Vanilla-QRL under the single abrupt Hamiltonian switching scenario. Compared with the stationary setting, the fidelity reward still increases steadily during training but converges to noticeably lower values across all system sizes. Furthermore, the performance degradation becomes more pronounced as the system size increases, indicating that identifying multiple Hamiltonians within a single trajectory is more challenging than learning a stationary Hamiltonian.

\begin{figure}[t]
    \centering

    \subfigure[2-qubit system]{
        \includegraphics[width=1\linewidth]{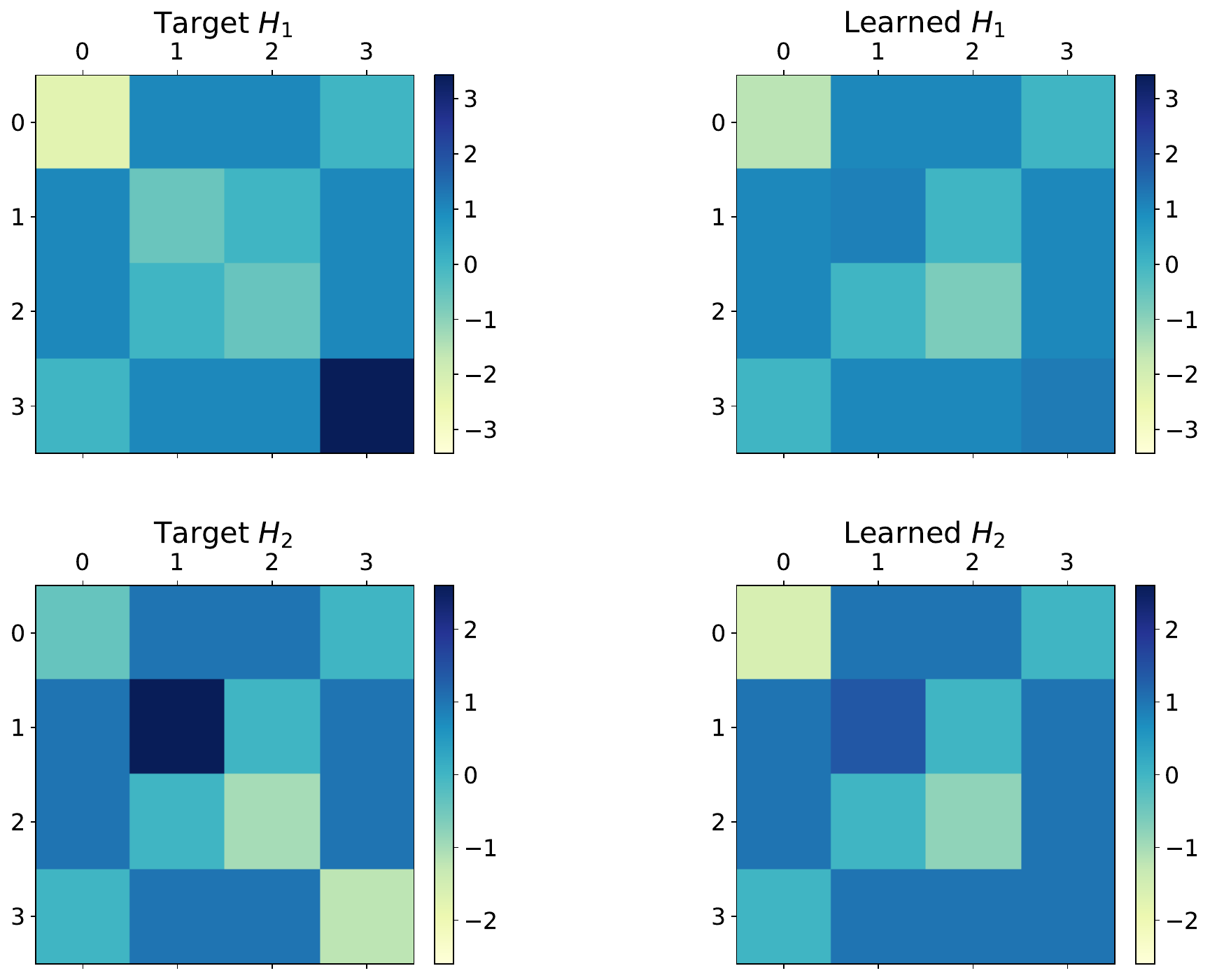}
        \label{fig:heatmap2_scen2_vanilla}
    }
    \hfill
    \subfigure[4-qubit system]{
        \includegraphics[width=1\linewidth]{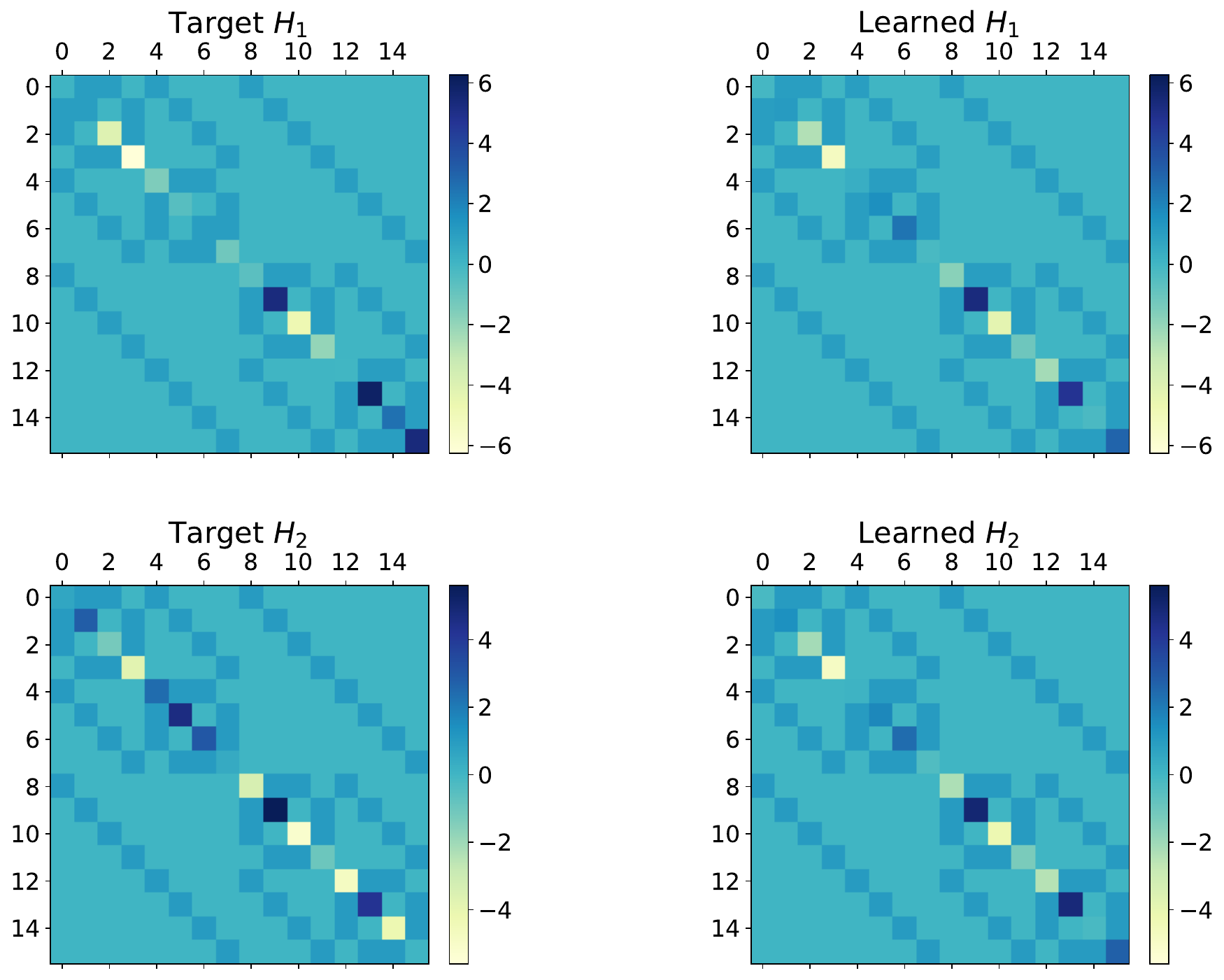}
        \label{fig:heatmap4_scen2_vanilla}
    }

    \caption{Representative target and learned Hamiltonian matrix heatmaps obtained using Vanilla-QRL under Scenario 2 (Single Abrupt Hamiltonian Switching).}
    \label{fig:heatmaps_scenario2_vanilla}
\end{figure}

Fig.~\ref{fig:heatmaps_scenario2_vanilla} compares representative target and learned Hamiltonian matrix heatmaps for the 2- and 4-qubit systems. Although the learned Hamiltonians preserve the overall structure of both target Hamiltonians, they exhibit larger deviations in the matrix elements than those observed in the stationary scenario, indicating that the abrupt Hamiltonian switch makes accurate reconstruction more challenging. These observations suggest that the feedforward Vanilla-QRL agent can partially recover the changing system dynamics, but struggles to accurately estimate both Hamiltonians from a single trajectory.

Table~\ref{tab:scenario2} quantitatively confirms these observations. The average fidelity gradually decreases as the system size increases, indicating that accurately reconstructing the changing system dynamics becomes more challenging for larger systems. Although the reported parameter MSE decreases with system size, the fidelity results demonstrate that the overall quality of the reconstructed quantum evolution deteriorates, suggesting that parameter errors alone do not fully capture the reconstruction performance in this non-stationary setting.

The reduced reconstruction accuracy observed in this scenario is expected, as the governing Hamiltonian changes during the trajectory while the Vanilla-QRL agent processes each observed quantum state without retaining information from previous time steps. Consequently, the agent cannot effectively exploit the temporal dependencies introduced by the Hamiltonian switch.

\subsubsection{GRU-QRL}
To address the limitations of the feedforward Vanilla-QRL agent under non-stationary dynamics, we replace the policy network with the GRU-QRL architecture. By incorporating recurrent memory, the GRU-QRL agent retains information from previously observed quantum states, enabling it to exploit the temporal dependencies introduced by the Hamiltonian switch. We evaluate the agent under the same single abrupt Hamiltonian switching scenario to assess whether recurrent memory improves Hamiltonian reconstruction.

\begin{figure}[h]
    \centering
    \includegraphics[width=1\linewidth]{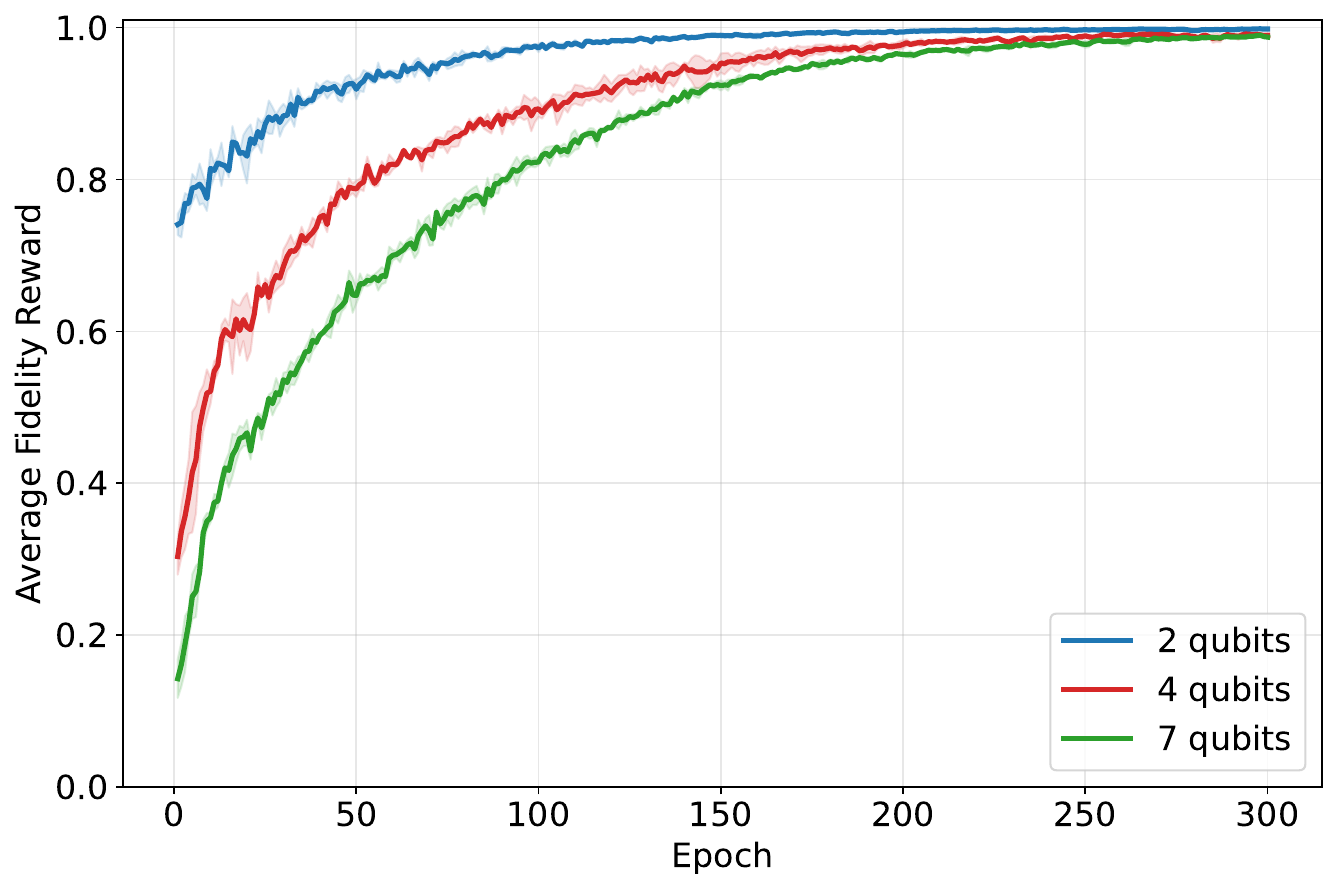}
    \caption{Learning curves of the GRU-QRL agent under Scenario 2 (Single Abrupt Hamiltonian Switching). Shaded regions indicate the standard deviation over three independent random seeds.}
    \label{fig:Scenario(2)_GRU_curve}
\end{figure}

Fig.~\ref{fig:Scenario(2)_GRU_curve} illustrates the learning curves of the proposed GRU-QRL agent for the 2-, 4-, and 7-qubit systems. In contrast to Vanilla-QRL, the recurrent agent converges rapidly to near-perfect fidelity across all system sizes, with stable training throughout the optimization process. Although convergence becomes slightly slower as the number of qubits increases, all systems achieve average fidelity rewards close to one by the end of training.

\begin{figure}[t]
    \centering

    \subfigure[2-qubit system]{
        \includegraphics[width=1\linewidth]{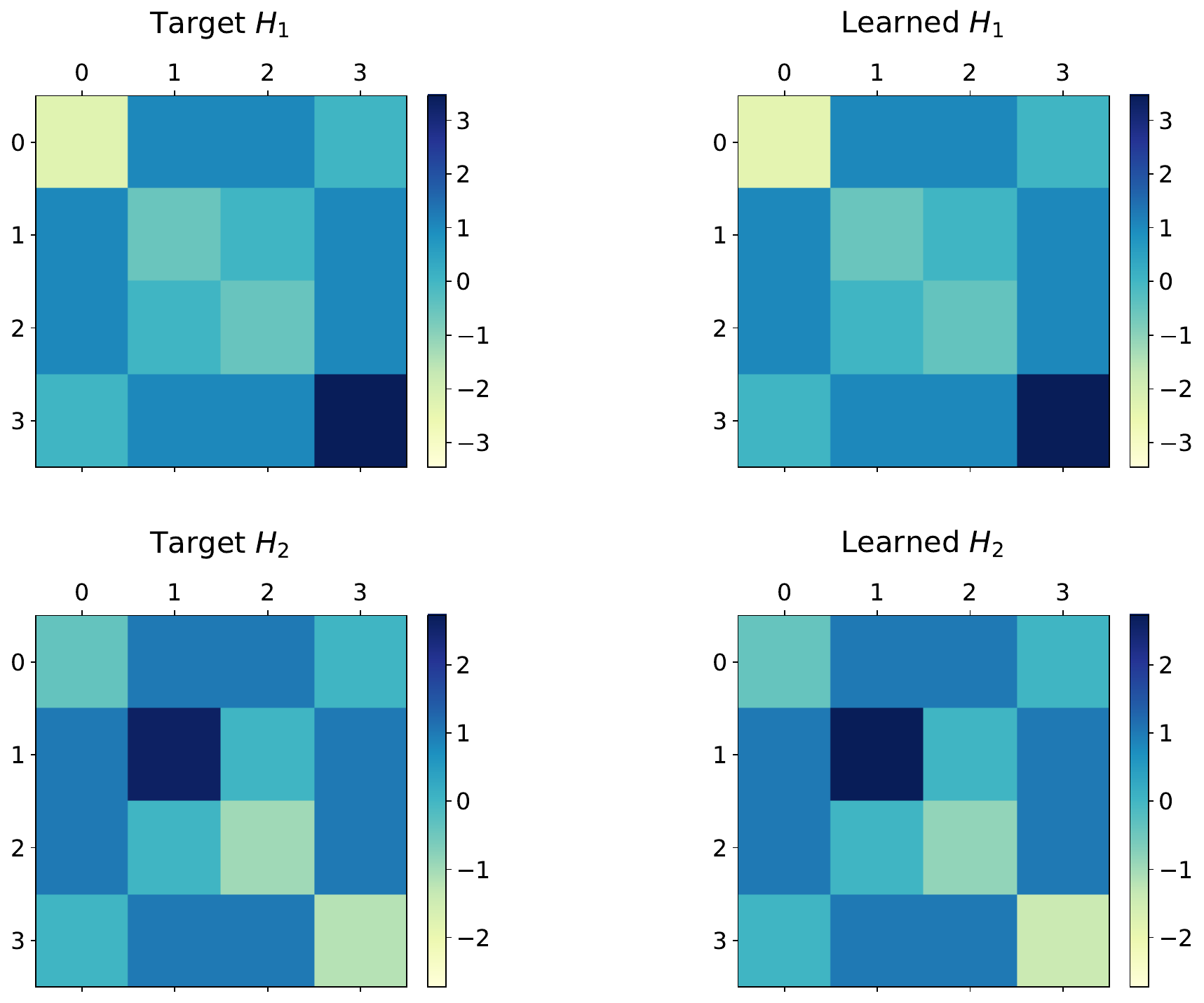}
        \label{fig:heatmap2_scen2_GRU}
    }
    \hfill
    \subfigure[4-qubit system]{
        \includegraphics[width=1\linewidth]{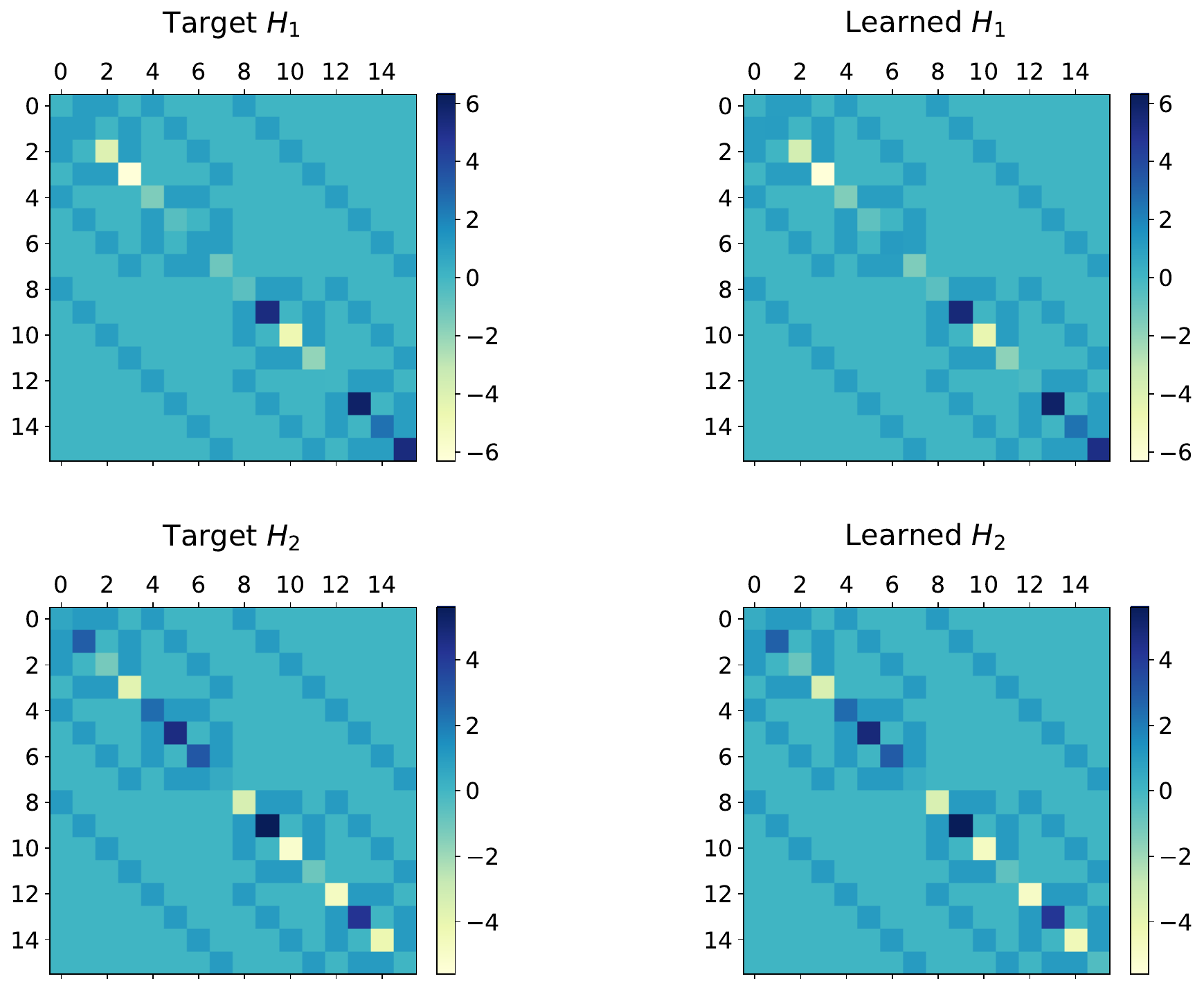}
        \label{fig:heatmap4_scen2_GRU}
    }

    \caption{Representative target and learned Hamiltonian matrix heatmaps obtained using GRU-QRL under Scenario 2 (Single Abrupt Hamiltonian Switching).}
    \label{fig:heatmaps_scenario2_GRU}
\end{figure}

Fig.~\ref{fig:heatmaps_scenario2_GRU} compares representative target and learned Hamiltonian matrix heatmaps for the 2- and 4-qubit systems. The learned Hamiltonians closely match the corresponding target Hamiltonians for both $H_1$ and $H_2$, accurately preserving the overall structure of the matrix elements. Compared with the Vanilla-QRL results, the reconstructed Hamiltonians exhibit substantially fewer reconstruction errors, demonstrating the effectiveness of recurrent memory in recovering the changing system dynamics.

Table~\ref{tab:scenario2} summarizes the quantitative performance of GRU-QRL under the single abrupt Hamiltonian switching scenario. The proposed recurrent architecture consistently achieves near-perfect reconstruction across all system sizes, with final fidelities exceeding $0.987$ and parameter MSEs below $0.012$ Compared with Vanilla-QRL, the incorporation of recurrent memory leads to substantial improvements in both fidelity and parameter estimation accuracy, confirming that exploiting temporal dependencies is essential for Hamiltonian learning under non-stationary dynamics.

QGRNN performs similarly to the feedforward Vanilla-QRL agent under the single abrupt switching scenario, with neither method able to match the reconstruction accuracy of GRU-QRL. As shown in Table~\ref{tab:scenario2}, QGRNN and Vanilla-QRL achieve comparable fidelity and parameter MSE across all system sizes, while GRU-QRL achieves better performance. These results indicate that although QGRNN continuously updates its Hamiltonian parameters after each transition, these updates are not sufficient to accurately recover both Hamiltonians within the trajectory.

Upon closer inspection of the QGRNN parameter estimates, we observed that QGRNN was able to react to the change in the governing Hamiltonian. Specifically, after the switch from $H_1$ to $H_2$, its parameter estimates began moving toward the parameters of $H_2$. However, this adaptation was relatively slow, and the remaining transitions after the switch were not sufficient for the estimates to converge toward $H_2$.

These results also highlight an important architectural tradeoff between the two approaches. QGRNN directly parameterizes the candidate Hamiltonian and provides highly accurate reconstruction under stationary dynamics. However, when the governing dynamics change during operation, adaptation relies on continued optimization of its Hamiltonian parameters. In contrast, GRU-QRL incorporates recurrent memory within the policy, allowing its estimates to depend on the sequence of previously observed states and providing a more effective mechanism for tracking changes in the underlying dynamics. This distinction becomes particularly relevant for quantum-system monitoring and adaptive identification, where the governing Hamiltonian may vary during operation.

The strong performance achieved under a single Hamiltonian switch demonstrates that recurrent memory effectively addresses the limitations of the feedforward architecture. We next evaluate the proposed GRU-QRL agent under more challenging non-stationary settings.

\subsection{Scenario 3: Every-Step Switching}

We next evaluate the proposed GRU-QRL framework under a more challenging non-stationary setting, where the governing Hamiltonian changes at every time step of the observed trajectory. Unlike the single-switch scenario, the agent must continuously infer a sequence of different Hamiltonians from the evolving quantum states, thus increasing the complexity of the learning task.  Although such rapid Hamiltonian changes are less common in practical quantum systems, this scenario serves as an extreme stress test of the proposed framework, evaluating its ability to adapt under highly dynamic system behavior.

\begin{figure}[h]
    \centering
    \includegraphics[width=1\linewidth]{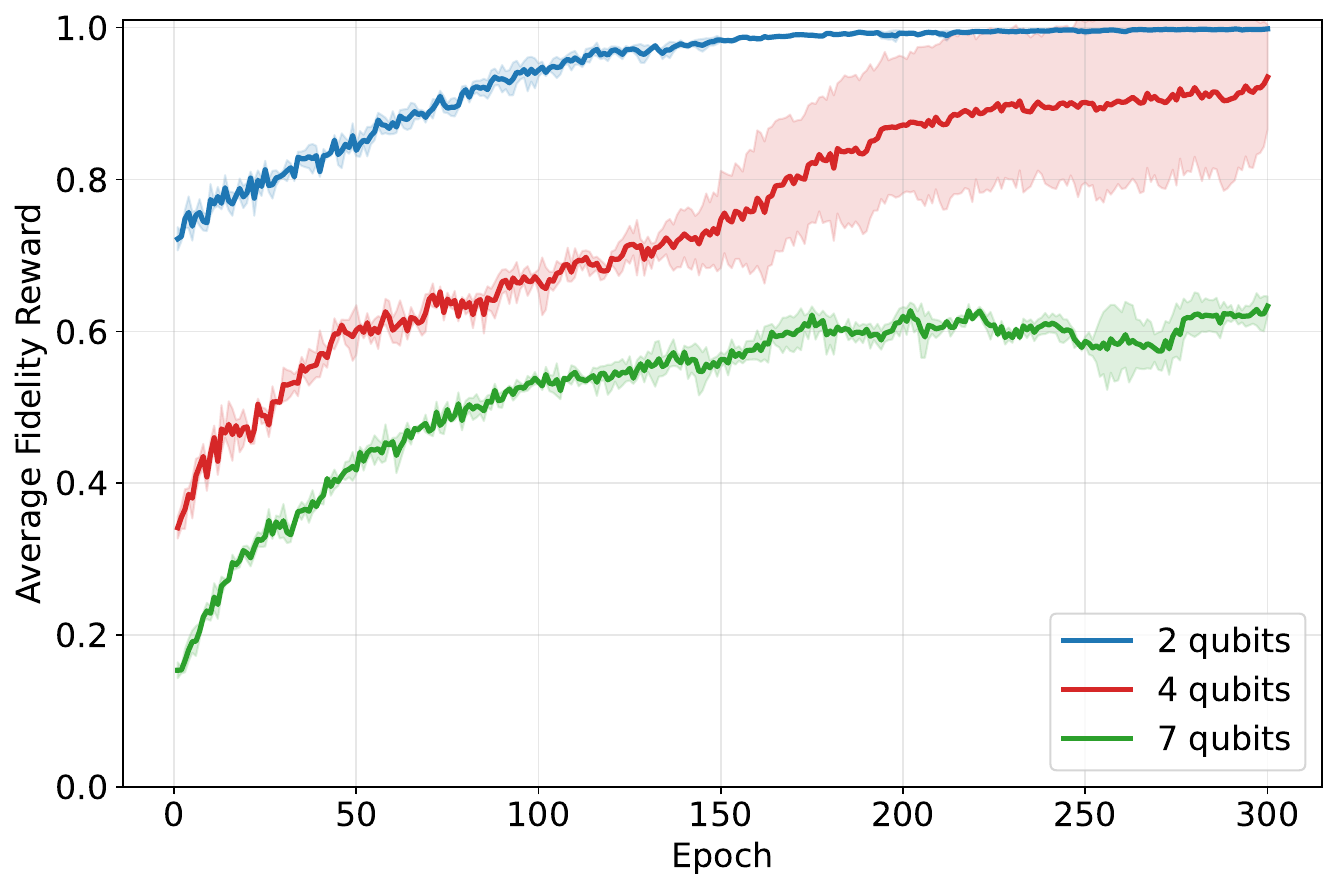}
    \caption{Learning curves of the GRU-QRL agent under Scenario 3 (every-step switching). Shaded regions indicate the standard deviation over three independent random seeds}
    \label{fig:Scenario(3)_curve}
\end{figure}

Fig.~\ref{fig:Scenario(3)_curve} presents the learning curves of the proposed GRU-QRL agent under continuously changing Hamiltonian dynamics. For the 2-qubit system, the agent converges rapidly to near-perfect fidelity. As the system size increases, however, convergence becomes slower and the final fidelity decreases, reflecting the increased difficulty of learning multiple time-varying Hamiltonians. While the 4-qubit system still achieves a high final fidelity, the 7-qubit system exhibits a noticeably lower plateau, indicating that this setting presents a substantially more challenging learning problem.

Table \ref{tab:scenario(3)} summarizes the quantitative performance of GRU-QRL under the continuously changing Hamiltonian scenario. The proposed framework maintains excellent reconstruction accuracy for the 2-qubit system, achieving a final fidelity of $0.998$ and a low parameter MSE. Performance gradually degrades for the 4- and 7-qubit systems, with the final fidelity decreasing to $0.935$ and $0.633$, respectively, while the parameter MSE increases accordingly. These results demonstrate that continuously changing Hamiltonians represent a considerably more challenging learning problem than the single-switch scenario, particularly as the system size increases.

QGRNN performs noticeably worse than GRU-QRL under the every-step switching scenario. As shown in Table~\ref{tab:scenario(3)}, the difference is particularly clear for the 2- and 4-qubit systems, where GRU-QRL maintains high fidelity while QGRNN experiences a decrease in performance. For the 7-qubit system, both approaches perform considerably worse, indicating that continuously identifying a different Hamiltonian at every time step becomes increasingly challenging as system size increases. The parameter MSE follows a similar trend, with QGRNN producing considerably larger reconstruction errors than GRU-QRL for the 2- and 4-qubit systems. Overall, these results show that GRU-QRL is better able to handle the rapidly changing Hamiltonian dynamics, although the performance degradation at 7 qubits suggests that this highly non-stationary setting remains challenging as the system size increases.

\begin{table*}[t]
\centering
\caption{Performance comparison of GRU-QRL and QGRNN under Scenario 3 (every-step switching). Results are reported as mean $\pm$ standard deviation over three independent random seeds.}
\label{tab:scenario(3)}
\setlength{\tabcolsep}{3pt}
\begin{tabular}{ccccc}
\hline
& \multicolumn{2}{c}{\textbf{GRU-QRL}}
& \multicolumn{2}{c}{\textbf{QGRNN}} \\
\cline{2-5}
\textbf{System Size}
& \textbf{Fidelity} & \textbf{MSE}
& \textbf{Fidelity} & \textbf{MSE} \\
\hline
2 qubits
& $0.998375 \pm 0.000505$
& $0.005308 \pm 0.004071$
& $0.866956 \pm 0.019904$
& $1.058368 \pm 0.020598$ \\

4 qubits
& $0.934933 \pm 0.069440$
& $0.133815 \pm 0.159934$
& $0.581043 \pm 0.015969$
& $1.080933 \pm 0.063460$ \\

7 qubits
& $0.633219 \pm 0.012978$
& $0.811587 \pm 0.039764$
& $0.525635 \pm 0.029117$
& $0.998313 \pm 0.028538$ \\
\hline
\end{tabular}
\end{table*}

\subsection{Scenario 4: Longer Trajectory with Random Abrupt Switching}

In this scenario, we extend the abrupt Hamiltonian switching setting in scenario 2 to a longer trajectory of $T=20$ time steps. Unlike Scenario 2, the switching point is no longer fixed and is randomly selected as $t_s \in \{5,10,15\}$ for each trajectory. This setting evaluates whether the agent can identify and adapt to an abrupt change in the governing Hamiltonian when the location of the switch is not known in advance.

\begin{figure}[h]
    \centering
    \includegraphics[width=1\linewidth]{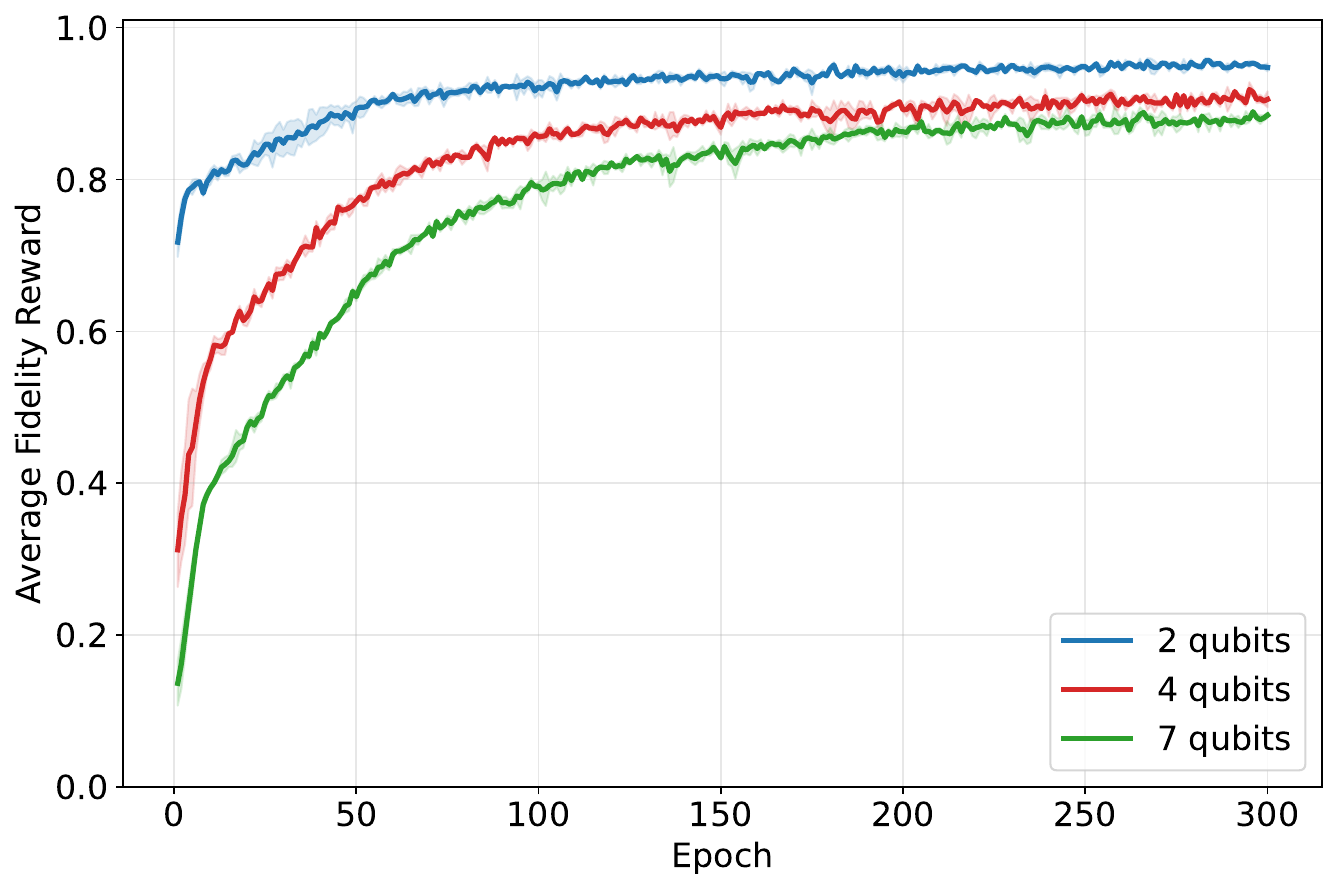}
    \caption{Learning curves of the GRU-QRL agent under Scenario 4 (Longer Trajectory with Random Abrupt Switching). Shaded regions indicate the standard deviation over three independent random seeds.}
    \label{fig:Scenario(4)_GRU_curve}
\end{figure}

Fig.~\ref{fig:Scenario(4)_GRU_curve} presents the learning curves of the GRU-QRL agent under the longer trajectory with random abrupt switching. The fidelity reward increases steadily throughout training for all system sizes, with the 2-qubit system converging faster and reaching the highest final fidelity. The 4- and 7-qubit systems converge more gradually, but both continue to improve throughout training and reach high fidelity by the final epochs. These results indicate that GRU-QRL is able to learn under longer trajectories even when the Hamiltonian switch occurs at different and unknown locations across trajectories. As the system size increases, however, convergence becomes slower and the final fidelity decreases slightly, reflecting the increased difficulty of adapting to the changing dynamics in larger systems.

\begin{figure}[t]
    \centering

    \subfigure[2-qubit system]{
        \includegraphics[width=1\linewidth]{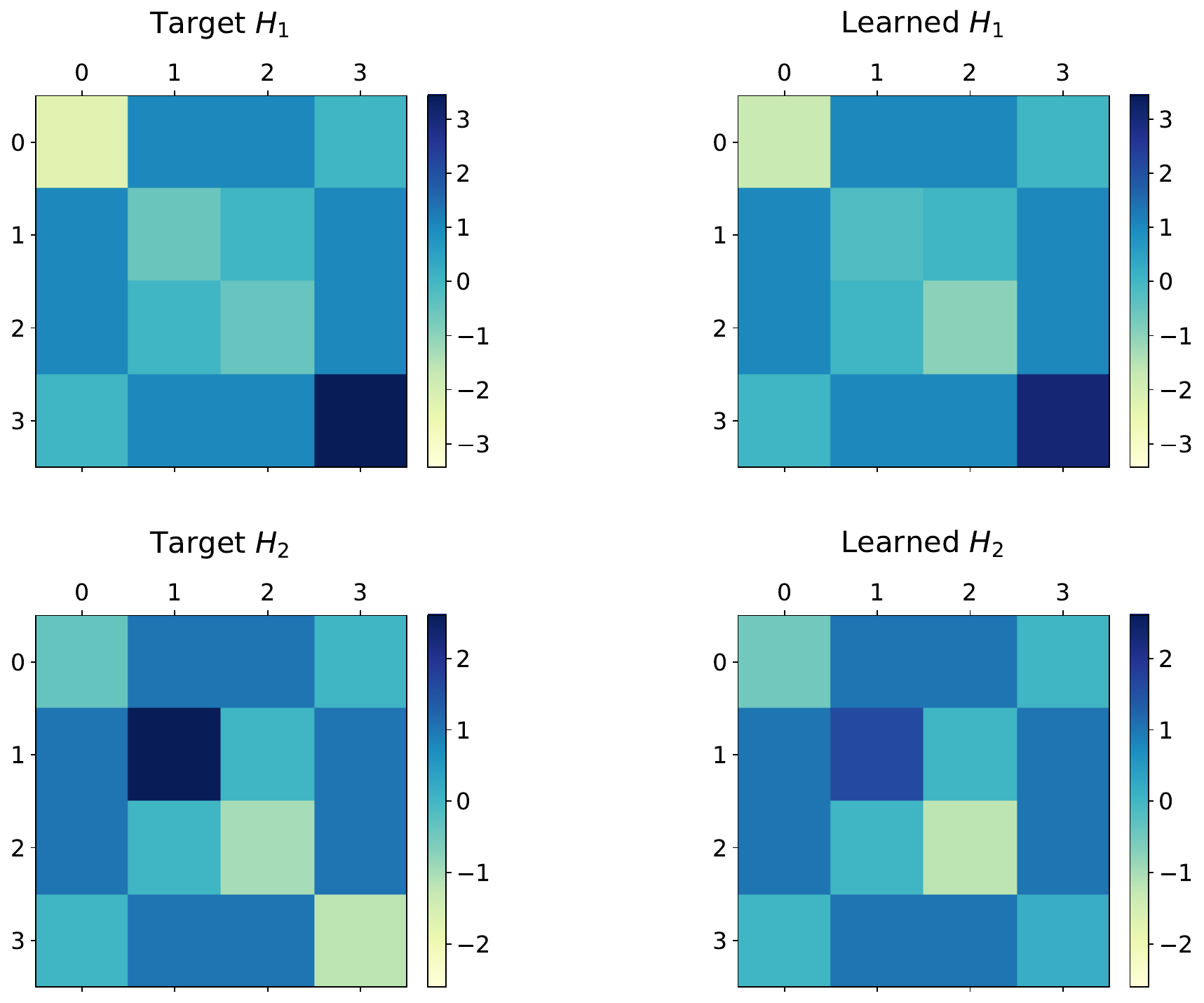}
        \label{fig:heatmap2_scen4_GRU}
    }
    \hfill
    \subfigure[4-qubit system]{
        \includegraphics[width=1\linewidth]{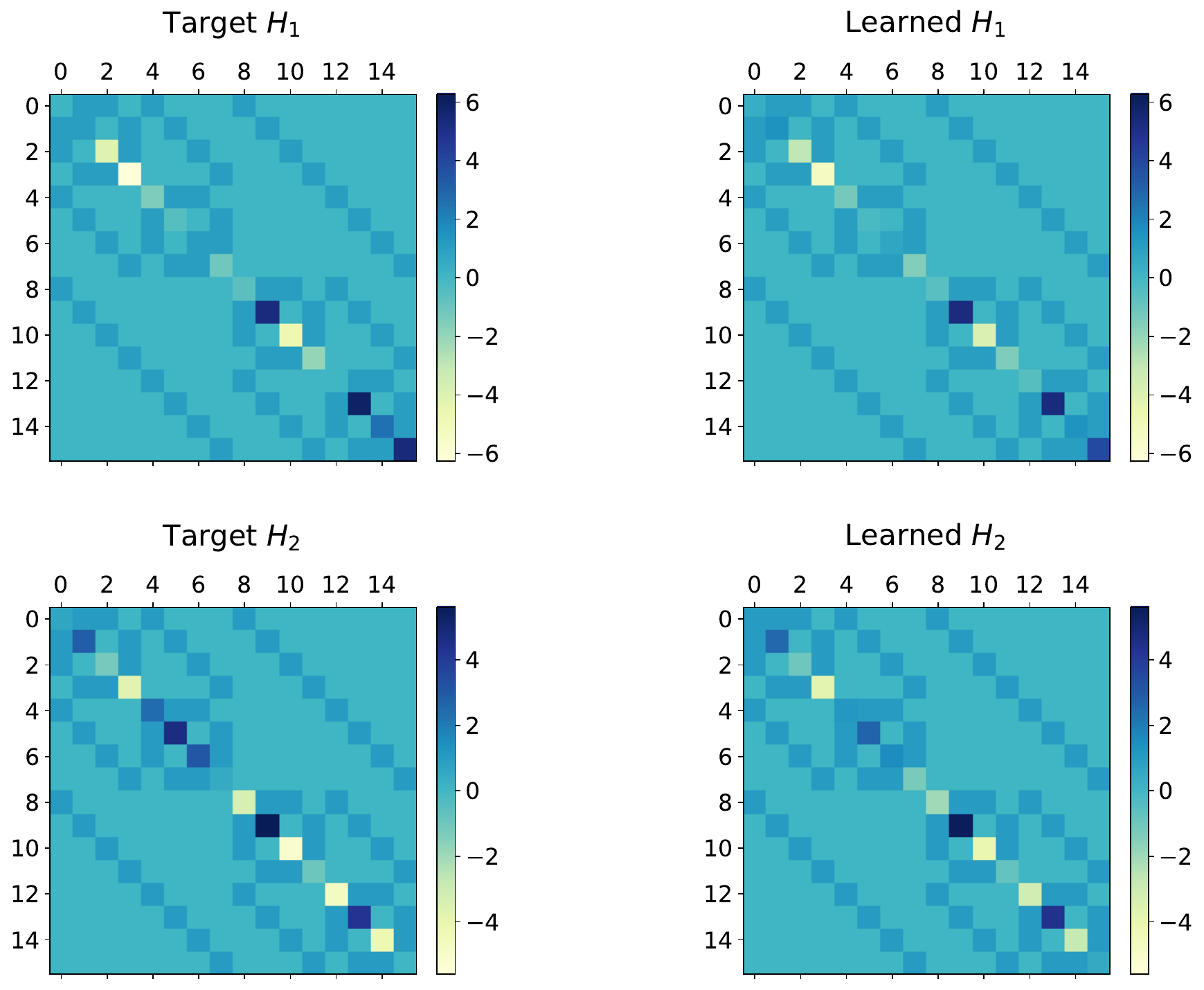}
        \label{fig:heatmap4_scen4_GRU}
    }

    \caption{Representative target and learned Hamiltonian matrix heatmaps obtained using GRU-QRL under Scenario 4 (Longer Trajectory with Random Abrupt Switching).}
    \label{fig:heatmaps_scenario4_GRU}
\end{figure}

Fig.~\ref{fig:heatmaps_scenario4_GRU} compares representative target and learned Hamiltonian matrix heatmaps for the 2- and 4-qubit systems under the longer trajectory with random abrupt switching. For both system sizes, the learned Hamiltonians closely resemble the corresponding target Hamiltonians for both $H_1$ and $H_2$.

Table~\ref{tab:scenario(4)} compares the performance of GRU-QRL and QGRNN under the longer trajectory with random abrupt switching. GRU-QRL consistently outperforms QGRNN across all system sizes, achieving higher fidelity and lower parameter MSE. The difference is particularly clear for the 2-qubit system, while the gap becomes smaller as the system size increases. Overall, these results show that GRU-QRL is better able to reconstruct the changing Hamiltonian dynamics when the switching location varies across trajectories.

\begin{table*}[t]
\centering
\caption{Performance comparison of GRU-QRL and QGRNN under Scenario 4 (Longer Trajectory with Random Abrupt Switching). Results are reported as mean $\pm$ standard deviation over three independent random seeds.}
\label{tab:scenario(4)}
\setlength{\tabcolsep}{3pt}
\begin{tabular}{ccccc}
\hline
& \multicolumn{2}{c}{\textbf{GRU-QRL}}
& \multicolumn{2}{c}{\textbf{QGRNN}} \\
\cline{2-5}
\textbf{System Size}
& \textbf{Fidelity} & \textbf{MSE}
& \textbf{Fidelity} & \textbf{MSE} \\
\hline
2 qubits
& $0.947584 \pm 0.005295$
& $0.158119 \pm 0.032667$
& $0.793782 \pm 0.072346$
& $1.523907 \pm 0.681913$ \\

4 qubits
& $0.905600 \pm 0.010573$
& $0.083598 \pm 0.014703$
& $0.699645 \pm 0.086112$
& $0.672778 \pm 0.241955$ \\

7 qubits
& $0.889234 \pm 0.003253$
& $0.040541 \pm 0.006636$
& $0.717961 \pm 0.077495$
& $0.392060 \pm 0.228401$ \\
\hline
\end{tabular}
\end{table*}

\begin{figure}[h]
    \centering
    \includegraphics[width=1\linewidth]{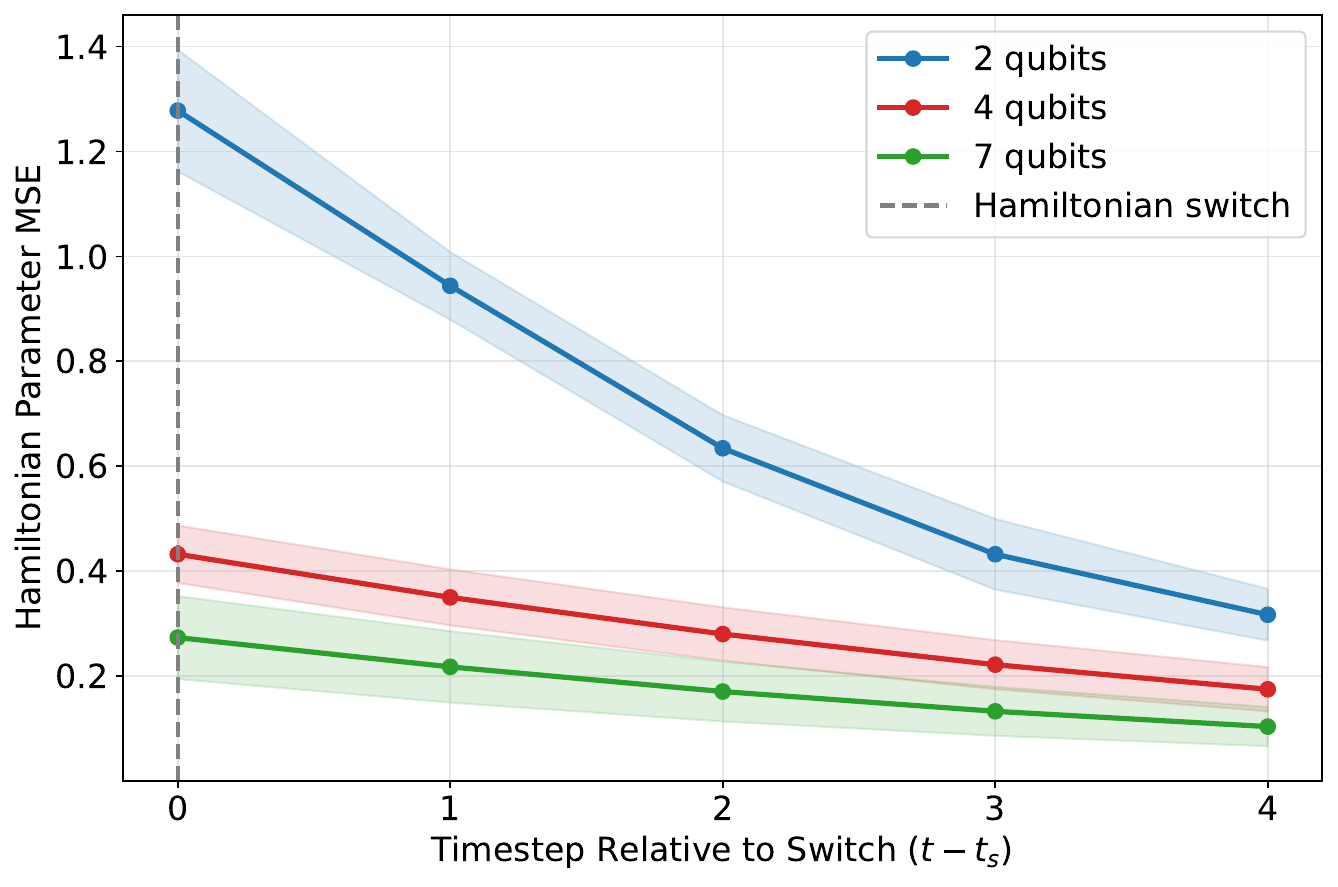}
    \caption{Post-switch adaptation of GRU-QRL under Scenario 4 (Longer Trajectory with Random Abrupt Switching). The parameter MSE is shown as a function of the time step relative to the Hamiltonian switch, with shaded regions indicating the standard deviation over three independent random seeds.}
    \label{fig:Scenario(4)_MSE_curve}
\end{figure}

Since the switching location varies randomly across trajectories and the final fidelity achieved by GRU-QRL is lower than in the previous scenarios for some system sizes, we further investigate how the agent adapts right after the Hamiltonian change, Fig.~\ref{fig:Scenario(4)_MSE_curve} shows the parameter MSE relative to the switching time step. Here, $t-t_s=0$ represents the first transition following the Hamiltonian switch, and the subsequent time steps show how the parameter estimates evolve after observing the new dynamics. The MSE decreases consistently after the switch across all system sizes, indicating that the agent progressively adjusts its parameter estimates toward the new Hamiltonian as additional post-switch states become available. The adaptation is particularly clear for the 2-qubit system, where the MSE decreases over the first few transitions following the switch. Although the initial error is smaller for the larger systems, the same decreasing trend is observed, showing that the agent is able to respond to the Hamiltonian change rather than simply maintaining its pre-switch estimate.

It is important to note that only the first five post-switch time steps are included in this analysis. This is because the latest possible switch occurs at $t_s=15$, leaving only five transitions after the switch in a trajectory of $T=20$. Therefore, the analysis is restricted to the first five post-switch time steps to ensure a consistent comparison across all possible switching locations $t_s \in \{5,10,15\}$.

\subsection{Scenario 5: Longer Trajectory with Continuous Gradual Hamiltonian Drift}

In this scenario, we consider a more gradual form of non-stationarity, where the Hamiltonian continuously transitions from $H_1$ to $H_2$ over a longer trajectory. This allows us to evaluate whether the agent can track gradual changes in the underlying system dynamics rather than responding to a sudden switch.

\begin{figure}[h]
    \centering
    \includegraphics[width=1\linewidth]{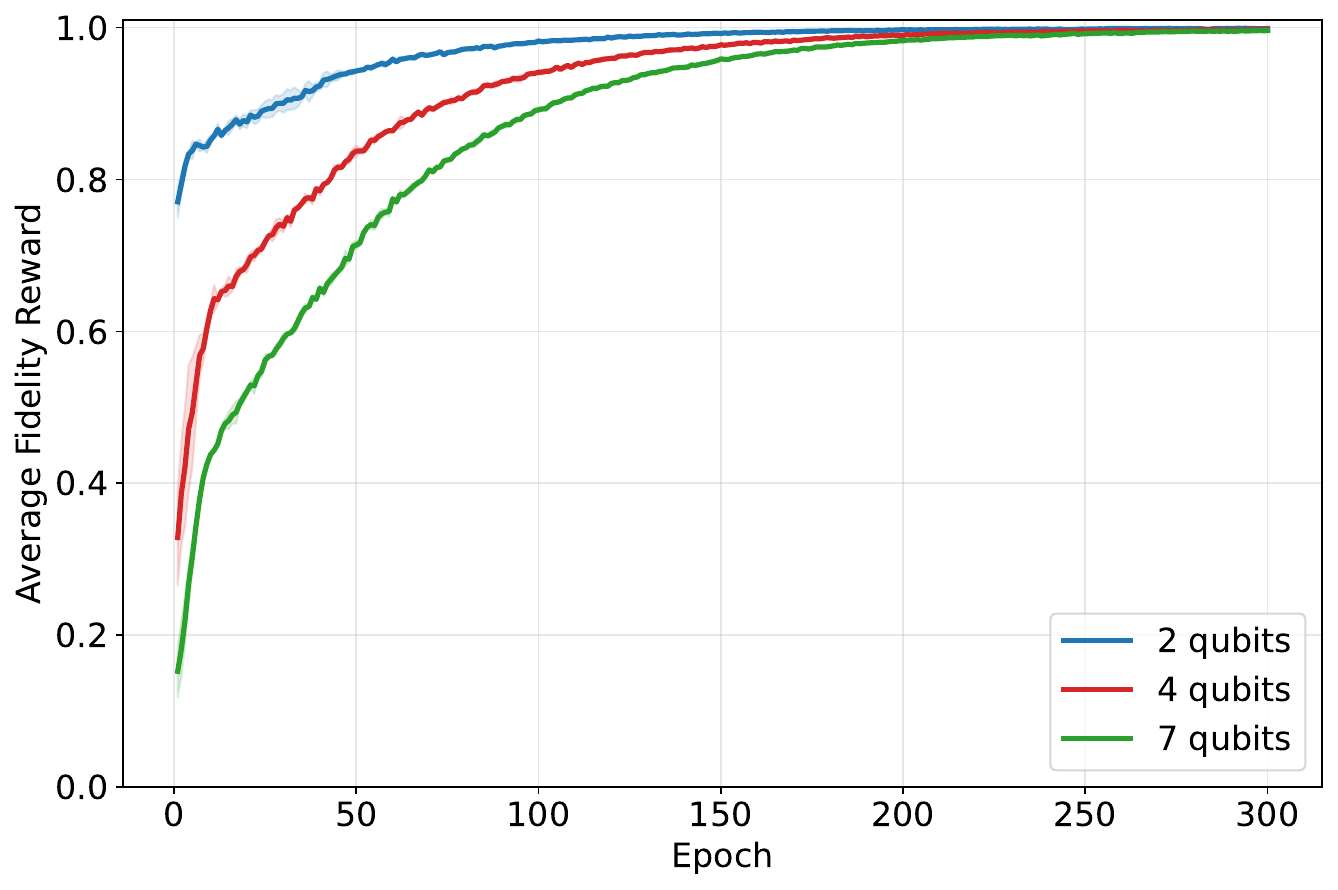}
    \caption{Learning curves of the GRU-QRL agent under Scenario 5 (Longer Trajectory with Continuous Gradual Hamiltonian Drift). Shaded regions indicate the standard deviation over three independent random seeds.}
    \label{fig:Scenario(5)_GRU_curve}
\end{figure}

Fig.~\ref{fig:Scenario(5)_GRU_curve} presents the learning curves of GRU-QRL under the continuous gradual Hamiltonian drift scenario. The fidelity reward increases steadily throughout training for all system sizes and eventually converges to values close to one. Compared with the random abrupt switching scenario, GRU-QRL achieves noticeably higher final fidelity despite the Hamiltonian changing at every time step. This can be attributed to the gradual nature of the change, where the Hamiltonian parameters vary only slightly between consecutive time steps, allowing the recurrent agent to continuously adjust its parameter estimates as the system progressively evolves from $H_1$ toward $H_2$. Although convergence becomes slower as the system size increases, all three systems eventually achieve near-perfect fidelity, demonstrating that GRU-QRL can effectively track continuously drifting Hamiltonian dynamics.

\begin{figure*}[t]
    \centering

    \subfigure[2-qubit system]{
        \includegraphics[width=0.47\textwidth]{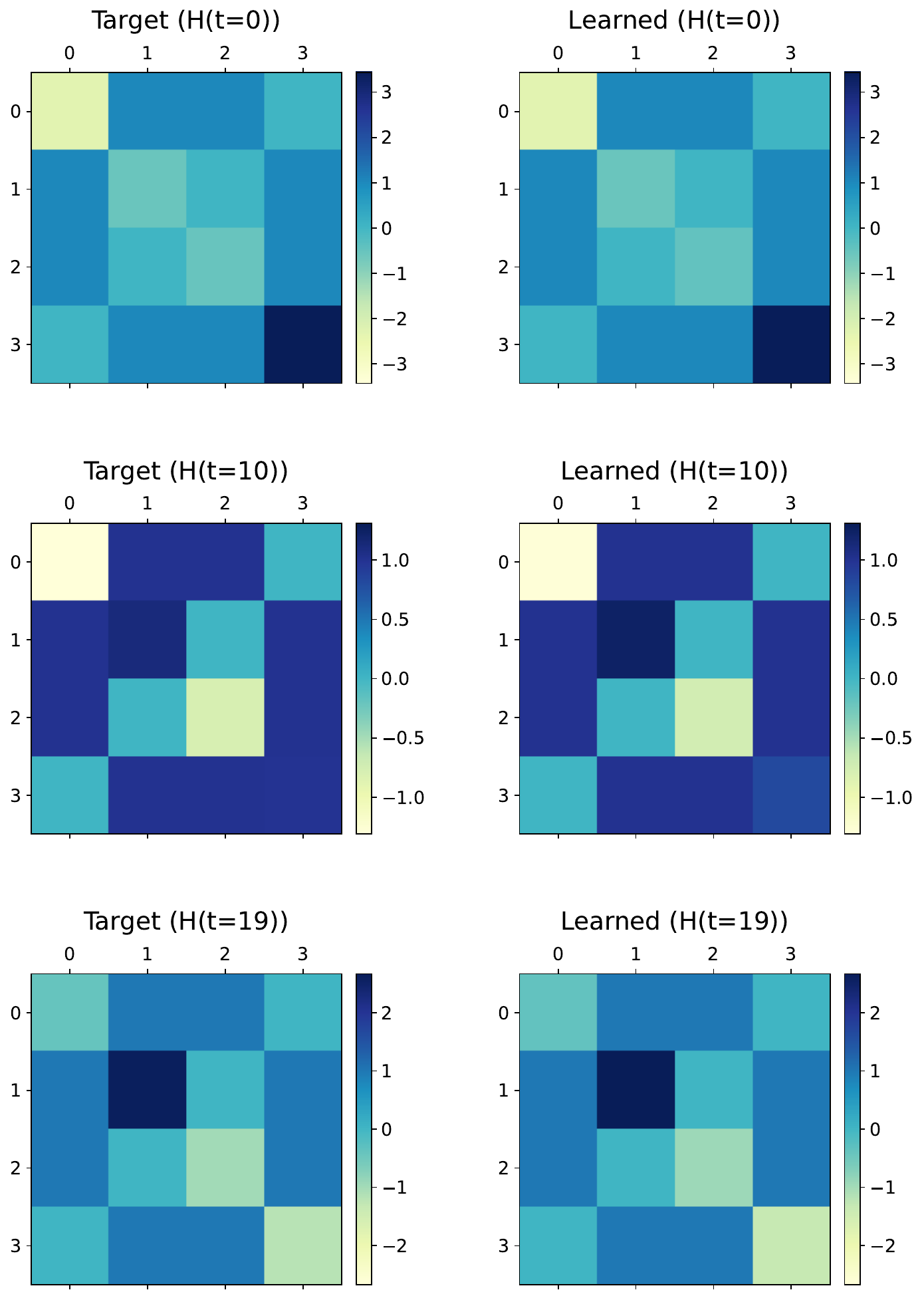}
        \label{fig:heatmap2_scen5_GRU}
    }
    \hspace{0.02\textwidth}
    \subfigure[4-qubit system]{
        \includegraphics[width=0.47\textwidth]{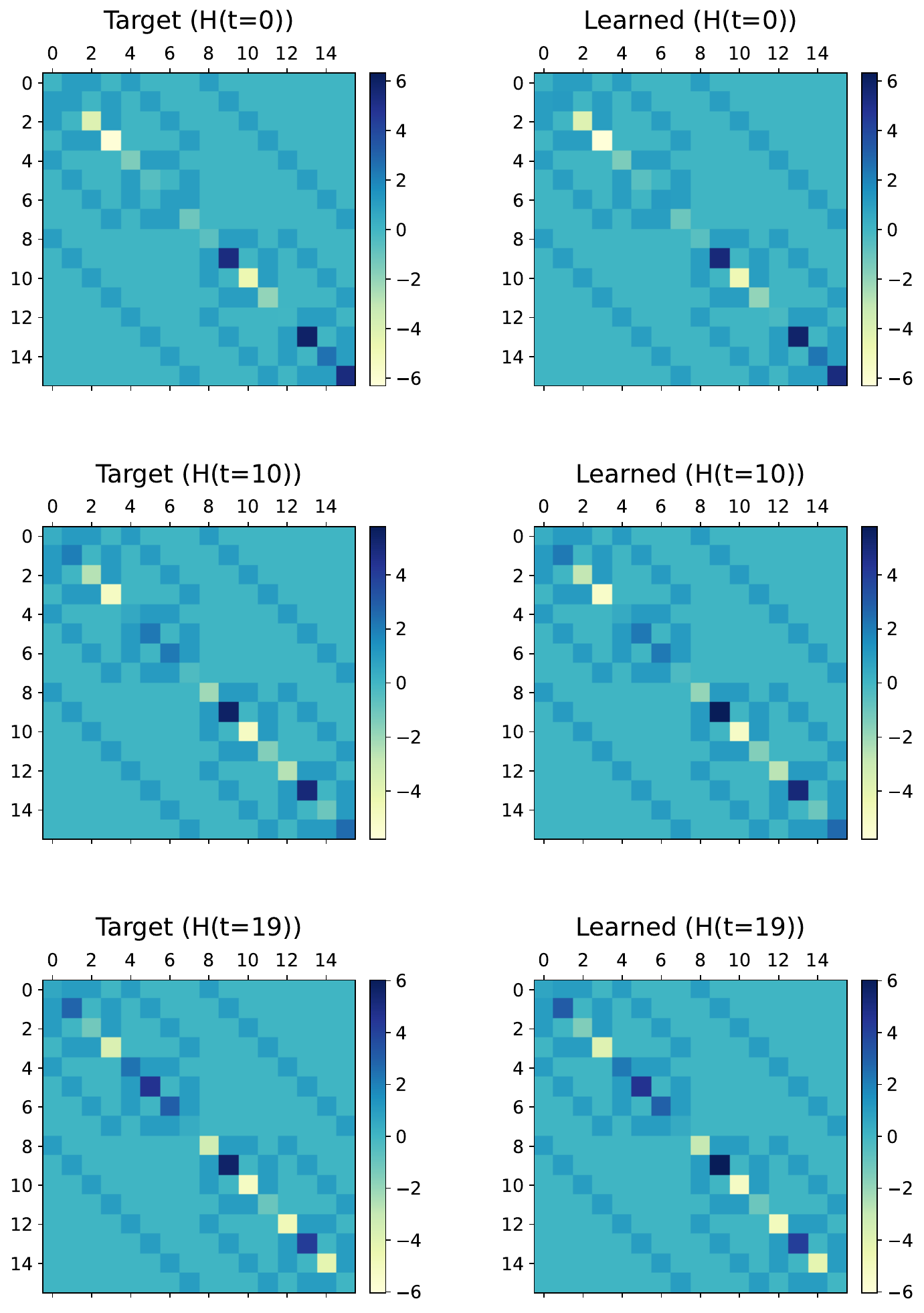}
        \label{fig:heatmap4_scen5_GRU}
    }

    \caption{Representative target and learned Hamiltonian matrix heatmaps obtained using GRU-QRL under Scenario 5 (Longer Trajectory with Continuous Gradual Hamiltonian Drift).}
    \label{fig:heatmaps_scenario5_GRU}
\end{figure*}

To further examine whether GRU-QRL is able to track the gradual change in the Hamiltonian throughout the trajectory, we compare the target and learned Hamiltonians at different stages of the evolution. Fig.~\ref{fig:heatmaps_scenario5_GRU} shows representative Hamiltonian matrix heatmaps at the beginning ($t=0$), middle ($t=10$), and end ($t=19$) of the trajectory for the 2- and 4-qubit systems. At all three points, the learned Hamiltonians closely match their corresponding target Hamiltonians, showing that the agent is not only able to recover $H_1$ and $H_2$, but also follows the intermediate Hamiltonian as the system gradually transitions between them.

Table~\ref{tab:scenario(5)} compares the performance of GRU-QRL and QGRNN under the continuous gradual Hamiltonian drift scenario. GRU-QRL substantially outperforms QGRNN across all system sizes, maintaining near-perfect fidelity and very low parameter MSE. In contrast, QGRNN achieves noticeably lower fidelity and higher parameter errors, indicating that it struggles to continuously track the gradually changing Hamiltonian. Interestingly, GRU-QRL performs particularly well under gradual drift, suggesting that smooth changes between consecutive Hamiltonians are easier for the agent to track than the abrupt changes considered in the previous scenarios.

\begin{table*}[t]
\centering
\caption{Performance comparison of GRU-QRL and QGRNN under Scenario 5 (Longer Trajectory with Continuous Gradual Hamiltonian Drift). Results are reported as mean $\pm$ standard deviation over three independent random seeds.}
\label{tab:scenario(5)}
\setlength{\tabcolsep}{3pt}
\begin{tabular}{ccccc}
\hline
& \multicolumn{2}{c}{\textbf{GRU-QRL}}
& \multicolumn{2}{c}{\textbf{QGRNN}} \\
\cline{2-5}
\textbf{System Size}
& \textbf{Fidelity} & \textbf{MSE}
& \textbf{Fidelity} & \textbf{MSE} \\
\hline
2 qubits
& $0.998932 \pm 0.000185$
& $0.003504 \pm 0.001209$
& $0.870578 \pm 0.077483$
& $1.888091 \pm 1.013442$ \\

4 qubits
& $0.997529 \pm 0.000364$
& $0.001917 \pm 0.000643$
& $0.760948 \pm 0.087958$
& $0.740193 \pm 0.261243$ \\

7 qubits
& $0.996209\pm 0.0002607$
& $0.001395 \pm 0.000248$
& $0.794595 \pm 0.012523$
& $0.428393 \pm 0.037390$ \\
\hline
\end{tabular}
\end{table*}

Overall, these results indicate that the proposed GRU-QRL framework effectively captures temporal dependencies under non-stationary quantum dynamics. Although reconstruction accuracy decreases as both the system size and the degree of non-stationarity increase, the agent remains capable of learning meaningful Hamiltonian representations without requiring prior knowledge of when the Hamiltonian changes.

\subsection{ Evaluation of Stationary and Non-Stationary Dynamics}

To further assess whether the proposed models overfit the training trajectories and evaluate their performance without fidelity-based reward feedback, we conduct additional evaluation experiments on unseen trajectories. The trained policies are frozen and evaluated without providing fidelity feedback reward to the agents. Here, fidelity is computed only as an evaluation metric to assess reconstruction accuracy. This allows us to assess their ability to reconstruct the governing Hamiltonian without further training or reward feedback.

For this experiment, we selected Scenario 1 and Scenario 4 as representative settings. Scenario 1 considers a fixed Hamiltonian throughout the trajectory, allowing us to evaluate the ability of Vanilla-QRL to reconstruct the governing Hamiltonian from unseen initial quantum states. In contrast, Scenario 4 considers a longer trajectory with a randomly located abrupt Hamiltonian switch, allowing us to evaluate the ability of GRU-QRL to adapt to changes in the system dynamics at different time steps.

Table~\ref{tab:evaluation} presents the training and evaluation results for Scenarios 1 and 4. Under Scenario 1, Vanilla-QRL maintains near-perfect fidelity and MSE across all system sizes on unseen trajectories. 

For Scenario 4, the switching times of the evaluation trajectories were randomly selected from the previously defined set. Similarly, under Scenario 4, GRU-QRL maintains comparable fidelity on unseen trajectories, although the evaluation MSE is slightly higher across all system sizes. These results suggest that both agents generalize to unseen initial quantum states without overfitting to the training trajectories. More importantly, the results for Scenario 4 suggest that GRU-QRL has learned to extract relevant dynamical information from the observed quantum states and use its recurrent memory to track changes in the underlying Hamiltonian, even when the switching time is randomly selected.

\begin{table*}[t]
\centering
\caption{Comparison of training and evaluation performance under stationary (Scenario 1) and non-stationary (Scenario 4) dynamics. Results are reported as mean $\pm$ standard deviation over three independent random seeds.}
\label{tab:evaluation}
\begin{tabular}{llcccc}
\hline
\textbf{Scenario} & \textbf{System Size}
& \textbf{Training Fidelity}
& \textbf{Evaluation Fidelity}
& \textbf{Training MSE}
& \textbf{Evaluation MSE} \\
\hline
\hline
\multirow{3}{*}{Scenario 1}
& 2 qubits & $0.999512 \pm 0.000098$ & $0.999894 \pm 0.000063$ & $0.000452 \pm 0.000188$ & $0.000691 \pm 0.000364$ \\
& 4 qubits & $0.997134 \pm 0.000499$ & $0.999104 \pm 0.000292$ & $0.001288 \pm 0.000799$& $0.002034 \pm 0.000644$ \\
& 7 qubits & $0.994119 \pm 0.000593$ & $0.994385 \pm 0.000217$ & $0.001222 \pm 0.000336$ & $0.001523 \pm 0.000262$ \\   
\hline
\multirow{3}{*}{Scenario 4}
& 2 qubits & $0.947584 \pm 0.005295$ & $0.951890 \pm 0.002425$ & $0.158119 \pm 0.032667$ & $0.202653 \pm 0.024053$ \\
& 4 qubits & $0.905600 \pm 0.010573$ & $0.915565 \pm 0.008228$ & $0.083598 \pm 0.014703$ & $0.107473 \pm 0.019676$ \\
& 7 qubits & $0.889234 \pm 0.003253$ & $0.900764 \pm 0.006010$ & $0.040541 \pm 0.006636$ & $0.068333 \pm 0.005406$ \\
\hline
\end{tabular}

\end{table*}

\section{Engineering and Implementation Considerations}\label{sec:engineering}

The proposed framework is intended as a step toward adaptive quantum-system identification in settings where the effective Hamiltonian may vary during operation. In our experiments, the QRL agent uses an $n$-qubit parameterized quantum circuit with three variational layers to target an $n$-qubit system. Each layer consists of parameterized single-qubit rotations and entangling operations, and the circuit outputs $n$ Pauli-$Z$ expectation values that are subsequently processed by either a linear readout or a GRU with hidden dimension 16.

\subsection{Architecture Considerations}
From an engineering perspective, the results reveal different operating characteristics for the considered Hamiltonian-learning architectures. In the evaluated settings, QGRNN provides highly accurate reconstruction under stationary dynamics, while GRU-QRL provides substantially better tracking when the Hamiltonian changes abruptly or gradually. This suggests that recurrent QRL may be particularly appropriate for monitoring and adaptive identification applications in which system parameters cannot be assumed to remain stationary throughout operation.

This work uses simulated closed-system dynamics and ideal quantum-state observations. In a physical implementation, expectation-value estimation would require repeated state preparation and finite-shot measurements, while state-preparation errors, decoherence, gate imperfections, and measurement noise may reduce reconstruction accuracy. Evaluating these effects experimentally, together with the measurement and computational overhead required for online operation, represents an important direction for future work.

\subsection{Scalability Considerations}
For the general $n$-qubit ring considered in this work, each variational layer contains $2n$ single-qubit rotation gates and $n$ CZ gates arranged in a ring. With three layers, the quantum policy therefore requires $6n$ parameterized rotations and $3n$ entangling gates, giving a total of $9n$ gates, excluding measurement and state preparation. Therefore, the number of trainable quantum parameters scales linearly with the system size. The number of trainable parameters in the classical components also scales linearly with $n$, as both the GRU input dimension and the output layer size increase with the system size while the hidden dimension remains fixed. However, a larger hidden dimension may be required for more complex systems.

A significant computational limitation of the present implementation arises from the classical simulation of the quantum environment. Generating state trajectories requires exponentiating Hamiltonian matrices of dimension $2^n \times 2^n$, while evaluating fidelity-based rewards involves computing overlaps between state vectors of dimension $2^n$. Therefore, the computational and memory requirements increase exponentially with the number of qubits, limiting the present experiments to $n \leq 7$. However, on quantum hardware, the time-evolved states could instead be obtained directly from the physical system. Moreover, the fidelity-based reward could be estimated by implementing the predicted evolution using Trotterization and comparing the resulting state with the true subsequent state through a SWAP-test-based fidelity estimation, at the cost of additional quantum resources. Within the investigated system sizes, the results show limited performance degradation under stationary and gradually drifting dynamics as the number of qubits increases. In contrast, performance decreases considerably under every-step Hamiltonian switching, indicating that rapidly varying dynamics become more challenging for larger systems. These results suggest that scalability depends not only on the system size but also on the rate at which the underlying Hamiltonian changes. It is worth noting that the same hyperparameter configuration was used across all system sizes. Further tuning of the circuit depth, learning rate, and GRU hidden dimension for larger systems may improve learning performance, particularly under rapidly changing Hamiltonian dynamics. Investigating these configurations could provide further insight into the scalability of the proposed framework.

\subsection{Hardware Outlook}
It is important to note that this work does not claim a distinct quantum computational advantage. Rather, the main contribution is an adaptive approach for identifying Hamiltonians under non-stationary dynamics. The relatively shallow quantum policy provides a potential starting point for implementation on NISQ devices. The policy circuit consists of single-qubit rotations and CZ gates, with measurements restricted to single-qubit Pauli-$Z$ expectation values, making it a relatively simple architecture for potential hardware implementation. For fault-tolerant quantum systems, the framework could potentially support continuous monitoring and recalibration of Hamiltonian parameters that drift during operation. A possible path toward post-classical computational capabilities lies in processing quantum-state information directly on quantum hardware, reducing reliance on explicit classical representations of large quantum states, which become increasingly difficult to simulate as the system size grows.

\section{Conclusion} \label{sec:conclusion}

In this paper, we propose a QRL framework for Hamiltonian learning from sequential quantum state observations. By formulating Hamiltonian learning as a sequential decision-making problem, the proposed framework enables an agent to infer unknown Hamiltonian parameters directly from quantum state trajectories. We first demonstrate that the proposed Vanilla-QRL agent accurately reconstructs stationary Hamiltonians and extend the framework with a recurrent GRU-QRL agent to address non-stationary quantum dynamics. Experimental results on systems of different sizes showed that incorporating recurrent memory significantly improves Hamiltonian reconstruction when the underlying Hamiltonian changes during the observed trajectory. 

GRU-QRL demonstrates strong performance across the considered non-stationary settings, effectively adapting to both abrupt and gradual changes in the governing Hamiltonian. The agent maintains high reconstruction accuracy across different system sizes under single and randomly located abrupt switches, while achieving near-perfect reconstruction under gradual Hamiltonian drift. Under the more extreme every-step switching scenario, which was introduced as a stress test, GRU-QRL maintains relatively strong performance for the 2- and 4-qubit systems, while its performance decreases considerably for the 7-qubit system, highlighting the increasing difficulty of tracking rapidly changing Hamiltonian dynamics as the system size increases.

Overall, the results demonstrate that incorporating recurrent memory into a QRL-based Hamiltonian identification framework provides an effective mechanism for tracking time-varying quantum dynamics from sequential quantum-state observations. These findings support the use of recurrent QRL as a promising approach for adaptive quantum-system identification and monitoring.



\bibliographystyle{IEEEtran}
\bibliography{ref.bib}
\end{document}